\pdfoutput=1

\documentclass[11pt,twoside,a4paper,cmspaper,final,collab]{cms-tdr}
\def\svnVersion{166526}\def\cmsCernNo{2012-085}\def\cmsCernDate{\today}\def\cmsMessage{Submitted to the European Physical Journal C}

\begin{document}\cmsNoteHeader{QCD-11-012}

\hyphenation{had-ron-i-za-tion}
\hyphenation{cal-or-i-me-ter}
\hyphenation{de-vices}

\RCS$Revision: 120229 $
\RCS$HeadURL: svn+ssh://svn.cern.ch/reps/tdr2/papers/QCD-11-012/trunk/QCD-11-012.tex $
\RCS$Id: QCD-11-012.tex 120229 2012-05-07 10:06:11Z sunil $
\cmsNoteHeader{QCD-11-012} 
\title{Measurement of the underlying event in the Drell--Yan process in proton-proton collisions at \texorpdfstring{$\sqrt{s} = 7\TeV$}{sqrt(s) = 7 TeV}}

\date{\today}

\abstract{
A measurement of the underlying event (UE) activity in proton-proton collisions at a centre-of-mass energy of 7\TeV is performed using Drell--Yan events in a data sample corresponding to an integrated luminosity of 2.2\fbinv, collected by the CMS experiment at the LHC.
 The activity measured in the muonic final state ($\mathrm{q\overline q} \to \mu^+\mu^-$) is corrected to the particle level and compared with the predictions of various Monte Carlo generators and hadronization models.
 The dependence of the UE activity  on the dimuon invariant mass is well described by \PYTHIA and \HERWIG{}++ tunes derived from the leading jet/track approach, illustrating the universality of the UE activity.
 The UE activity is observed to be independent of the dimuon invariant mass in the region above 40\GeVcc, while a slow increase is observed with increasing transverse momentum of the dimuon system.
 The dependence of the UE activity on the transverse momentum of the dimuon system is accurately described by \MADGRAPH,  which simulates multiple hard emissions.
}

\hypersetup{%
pdfauthor={CMS Collaboration},%
pdftitle={Measurement of the underlying event in the Drell-Yan process in proton-proton collisions at sqrt(s) = 7 TeV},%
pdfsubject={CMS},%
pdfkeywords={CMS, physics, underlying event, Drell-Yan}}

\maketitle 

\section{Introduction}
In hadron-hadron scattering, the ``underlying event'' (UE) is defined as any hadronic activity that cannot be attributed to the particles originating from the hard scattering, which is characterized by a large momentum transfer, or to the hadronization of initial- and final-state radiation.
The UE activity is thus due to the hadronization of partonic constituents, not involved in the hard scattering, that have undergone multiple-parton interactions (MPIs) and to the hadronization of beam remnants that did not participate in other scatterings.
 These semihard interactions cannot be completely described by perturbative quantum chromodynamics (QCD) and require a phenomenological description involving parameters that must be tuned with the help of data~\cite{Bartalini:2010su}.

 The experimental study of the UE probes various aspects of hadron production in high energy hadron-hadron collisions.
 In particular it is sensitive to the interplay of perturbative methods describing the hard process and phenomenological models of the soft interactions that attempt to simultaneously describe MPIs, initial- and final-state radiation, the colour flow between final state partons, and the hadronisation process.
  Understanding the UE in terms of particle and energy densities will lead to better modelling by Monte Carlo programs that are used in precise measurements of standard model processes and searches for new physics at high energies.
 The UE affects the estimation of the efficiency of isolation criteria applied to photons and charged leptons, and the energy scale in jet identification.
 It also affects the reconstruction efficiency for processes like H$\rightarrow\gamma\gamma$, where the primary vertex is partly determined from the charged particles originating from the UE.
Hard MPIs are an important background for new physics searches, e.g. same-sign W production from MPIs~\cite{sameW} is a possible background to the same-sign double lepton SUSY searches~\cite{SUSY}.

The Compact Muon Solenoid (CMS)~\cite{CMS_JINST}, ATLAS, and ALICE experiments have carried out UE measurements at centre-of-mass energies ($\sqrt{s}$) of 0.9\TeV and 7\TeV using hadronic events (minimum-bias and single-jet triggered) containing a leading track-jet~\cite{CMS-PAS-QCD-10-001, CMS-PAS-QCD-10-010} or a leading track~\cite{atlasUE, aliceUE}.
 The analysis of the central charged particles and forward energy flow correlations in hard processes, e.g. $\rm pp \rightarrow$ W(\Z)X $\rightarrow \ell\nu(\ell\ell)$X~\cite{WZue}, provides supplementary insights into the nature of MPIs.
 In this paper, we use the Drell--Yan (DY) process~\cite{DY} with the muonic final state at a centre-of-mass energy of 7\TeV to perform a complementary UE measurement.
 The DY process with muonic final state is experimentally clean and theoretically well understood, allowing the particles from the UE to be reliably identified.
 The absence of QCD final-state radiation (FSR) permits a study of different kinematic regions with varying transverse momentum of $\gamma^{*}$/\Z  due to harder or softer initial-state radiation (ISR).
 The comparison of the UE measurement in DY events with QCD events having a leading track-jet is useful for probing the UE activity in different processes. UE measurements using the DY process have been reported previously in proton-antiproton collisions at $\sqrt{s}$ = 1.96\TeV~\cite{CDF:2010}.

The UE activity at a given centre-of-mass energy is expected to increase with the momentum transfer of the interaction.
 Events with a harder scale are expected to correspond, on average, to interactions with a smaller impact parameter and, in some models, to more MPIs~\cite{Sjostrand:1986ep, Frankfurt:2011}.
 This increased activity is observed to reach a plateau for high energy scales corresponding to small impact parameter.
 In this paper we investigate some aspects of the UE modelling in detail by measuring the invariant mass dependence of the UE activity for DY events with small transverse momentum of the DY system.
 This measurement separates the scale dependence of the UE activity from the ISR effect.
 The universality of the model parameters, denoted as tunes, implemented in the various MC programs is tested by comparing their predictions with our measurements.
 The portability of the UE parameters across different event generators, combined in some cases with different parton distribution functions (PDFs), is investigated as well.
 The modelling of the ISR is studied by measuring the UE activity as a function of the transverse momentum of the DY system.
 Finally, the dependence of the UE activity on ISR and FSR is determined by comparing the measurements from DY events with previous results from hadronic events containing a leading jet where FSR also plays a role.

The outline of the paper is as follows.
 Section 2 describes the various observables used in the present study.
 Section 3 summarizes the different MC models used and corresponding UE parameters.
 Section 4 presents experimental details: a brief detector  description, data samples, event and track selection criteria, correction procedure, and systematic uncertainties.
 Section~5 presents the results on UE activity measured in DY events and the comparison with the measurements based on a leading track-jet. The main results are summarized in Section 6.

\section{Observables}
The UE activity is measured in terms of  particle and energy densities.
 The particle density ($1/[\Delta\eta \Delta(\Delta\phi)] \langle N_{\rm ch} \rangle$) is computed as the average number of primary charged particles per unit pseudorapidity $\eta$ and per unit azimuthal separation $\Delta\phi$ (in radians) between a track and the transverse momentum of the dimuon system.
 The pseudorapidity is defined as $\eta = -\ln (\tan(\theta / 2))$, where $\theta$ is the polar angle measured with respect to the anticlockwise beam direction.
 The azimuthal angle $\phi$ is measured in the plane perpendicular to the beam axis.
 The energy density ($1/[\Delta\eta \Delta(\Delta\phi)] \langle \Sigma p_{T} \rangle$) is expressed in terms of the  average of the scalar sum of the transverse momenta of primary charged particles per unit pseudorapidity per unit azimuthal separation.
 The ratio of the energy and particle densities, as well as the total charged-particle multiplicity $N_{\rm ch}$ and the transverse momentum spectrum are also computed.
 The charged-particle multiplicity and transverse momentum distributions are normalized to unit area and to the average number of charged particles per event, respectively.
 Particles are considered as primary if they originate from the initial proton-proton interaction and are not the decay products of long-lived hadrons with a lifetime exceeding 10$^{-10}$\unit{s}.
 Apart from the muons from the DY process, all charged particles in the central region of the detector with pseudorapidity $|\eta| < 2$ and  with transverse momentum $p_{T} > 0.5$\GeVc are considered.

 The spatial distribution of the tracks is categorized by the azimuthal separation $\Delta\phi$.
 Particle production in the {\it away} region ($|\Delta\phi| > 120^{\circ}$) is expected to be dominated by the hardest ISR emissions, which balance the dimuon system.
 The {\it transverse} region ($60^{\circ} < |\Delta\phi| < 120^{\circ}$) and {\it towards} region ($|\Delta\phi| < 60^{\circ}$) are more sensitive to soft emissions and, in particular, those due to MPIs.
 The relevant information about the hard and the soft processes is extracted from the tracking and the muon systems of the CMS detector and thus the derived observables are insensitive to the uncertainties of the calorimetric measurements.
 The DY events with dimuon mass $M_{\mu\mu}$ around the \Z resonance are the least contaminated by background processes (heavy-quark, \cPqt\cPaqt, W+jets, and DY $\to \tau\tau$ production)~\cite{CMS_dy, CMS_wz} and best suited for the measurement of the UE activity.

 The UE activity is studied as a function of the magnitude of the dimuon transverse momentum ($p_{T}^{\mu\mu}=\abs{ \vec{p}^{\mu}_{T,1}+\vec{p}^{\mu}_{T,2}}$) and as a function of $M_{\mu\mu}$.
 The dependence  of the UE activity on $p_{T}^{\mu\mu}$ for high-mass dimuon pairs effectively probes the ISR spectrum.
 In order to minimize the background  contamination, the $p_{T}^{\mu\mu}$ dependence is studied only in the narrow mass window  $81 < M_{\mu\mu} < 101$\GeVcc.
 In contrast to the study of the UE activity in hadronic events using a leading track-jet~\cite{CMS-PAS-QCD-10-001, CMS-PAS-QCD-10-010}, this energy scale is sufficiently large to saturate the  MPI contributions.
 This observation is verified by studying the UE activity as a function of the dimuon mass in a wider mass range, where the total transverse momentum of the dimuon system is kept to a minimum by requiring $p_{T}^{\mu\mu} < 5\GeVc$.

\section{Monte Carlo models}
The UE dynamics are studied through the comparison of the observables in data with various tunes of 
\PYTHIA{}6~\cite{Sjostrand:2006za} and its successor \PYTHIA{}8~\cite{P8, Corke:2009pm}. \MADGRAPH (version 5)~\cite{Maltoni:2003, Alwall:2011}, which simulates up to six final-state fermions (including the muons), and \POWHEG~\cite{Frixione:2007}, which includes next-to-leading-order corrections on the hardest emission, are also compared to our measurements.
  For these two generators, softer emissions are simulated by $p_{T}$-ordered parton showers using \PYTHIA{}6 tunes and matched with the hard process produced by the generators.
 Hadronization in \PYTHIA{}6 and \PYTHIA{}8 is based on the Lund string fragmentation model~\cite{stringM}.
 The measurements are also compared to predictions of the \HERWIG{}++~\cite{hpp} angular-ordered parton shower and cluster hadronization model~\cite{hpp_hadron, hpp_hadron1}.

 The UE contributions from MPIs rely on modelling and tuning of the parameters in the MC generators.
 The MPI model of PYTHIA relies on two fundamental assumptions~\cite{Sjostrand:1986ep}:
\begin{itemize}
\item The ratio of the 2$\rightarrow$2 partonic cross section, integrated above a transverse momentum cutoff scale, and the total of the hadronic cross section is a measure of the amount of MPIs.
The cutoff scale $p_{0T}$ is introduced to regularize an otherwise diverging partonic cross section,

\begin{equation}
\sigma(p_{T}) = \sigma(p_{0T})\frac{p_{T}^{4}}{(p_{T}^{2} + p_{0T}^{2})^{2}}\ ,
\end{equation}

with

\begin{equation}
p_{0T} (\sqrt{s}) = p_{0T} (\sqrt{s_{0}})
\left( \frac{\sqrt{s}}{\sqrt{s_{0}}} \right)^{\epsilon} \ .
\end{equation}

Here $\sqrt{s_{0}} = 1.8\TeV$ and $\epsilon$ is a parameter characterizing the energy dependence of the cutoff scale.

\item The number of MPIs in an event has a Poisson distribution with a mean that depends on the overlap of the  matter distribution of the hadrons in impact-parameter space.
\end{itemize}

The MPI model used here~\cite{Skands:2007zg} includes showering of the MPI process, which  is interleaved with the ISR.

The tunes of the models vary mainly in the MPI regularization parameters, $p_{0T}$ and $\epsilon$, in the amount of colour reconnection, and in the PDF used.
 The Z1 tune~\cite{Z1} of \PYTHIA{}6 adopts the results of a global tuning performed by the ATLAS Collaboration~\cite{Buckley:2009bj} and uses the fragmentation and colour reconnection parameters of the  ATLAS AMBT1 tune~\cite{:2010ir}.
 The parameters of the Z1 tune related to the MPI regularization cutoff and its energy dependence are adjusted to describe previous CMS measurements of the UE activity in hadronic events~\cite{CMS-PAS-QCD-10-010} and uses the {CTEQ5L} PDF.
 The Z2 tune of \PYTHIA{}6 is an update of the Z1 tune using {CTEQ6L1}~\cite{Pumplin:2002vw}, the default used in most CMS generators; the regularization cutoff value at the nominal energy of $\sqrt{s_{0}}$ = 1.8\TeV is optimized to 1.832\GeVc.
The value of the energy evolution parameter for the Z2 tune is 0.275, as for the Z1 tune.
 The 4C~\cite{Corke:2010yf} tune of \PYTHIA{}8  follows a similar procedure as the ATLAS AMBT1 tune, but includes ALICE multiplicity data as well.
The values of the $p_{0T}(\sqrt{s_0})$ and $\epsilon$ parameters for the 4C tune are 2.085\GeVc and 0.19, respectively.
The effective value of $p_{0T}$ at $\sqrt{s} = 7\TeV$ is about 2.7\GeVc for both the Z2 and 4C tunes.

The LHC-UE7-2 tune of \HERWIG{}++ is based on ATLAS measurements of the UE activity in hadronic events~\cite{atlasUE}.
 The regularization cutoff parameter $p_{0T}$ for the LHC-UE7-2 tune is 3.36\GeVc at $\sqrt{s}$ = 7\TeV.
 The {CTEQ6L1} PDF is used in conjunction with \PYTHIA{}6 Z2, \PYTHIA{}8 4C, \MADGRAPH Z2, and \HERWIG{}++ LHC-UE7-2, while CT10~\cite{CT10} is used for \POWHEG, and {CTEQ5L} for the \PYTHIA{}6 Z1 simulations.

A comparison of these models with the measurements is presented in Section 5.

\section{Experimental methods}

The present analysis is performed with a sample of proton-proton collisions corresponding to an integrated luminosity of 2.2\fbinv, collected in March--August 2011 using the CMS detector~\cite{CMS_JINST}.

Muons are measured in the pseudorapidity range $|\eta| < 2.4$ with a detection system consisting of three subsystems: Drift Tubes, Cathode Strip Chambers, and Resistive Plate Chambers.
 Matching track segments from the muon detector to the tracks measured in the inner tracker results in a transverse momentum resolution between 1\% and 5\% for $p_T$ values up to 1\TeVc.
 The tracker subsystem consists of 1440 silicon-pixel and 15\,148 silicon-strip detector modules, and it measures charged particle trajectories within the nominal pseudorapidity range $|\eta| < 2.5$.
 The tracker is designed to provide a transverse impact parameter resolution of about 100\mum and a transverse momentum resolution of about 0.7\% for 1\GeVc charged particles at normal incidence ($\eta$ = 0).

The detector response is simulated in detail using the GEANT4 package~\cite{Agostinelli:2002hh}.
 The simulated signal and background events, including heavy-quark, \cPqt\cPaqt, W+jets, and DY $\to \tau\tau$ production, are processed and reconstructed in the same manner as collision data.

\subsection{Event and track selection}
The trigger requires the presence of at least two muon candidates.
 In periods of lower instantaneous luminosity both muons were required to have $p_{T} > 7\GeVc$, while in other periods the transverse momentum requirements were 13\GeVc and 8\GeVc for the leading and subleading muons, respectively.
 The trigger efficiency is above 95\% for the offline selected DY events with the requirement of $81 < M_{\mu\mu} < 101$\GeVcc.
 The offline selection requires exactly two muons reconstructed in the muon detector and the silicon tracker.
 Muon candidates are required to satisfy identification criteria based on the number of hits in the muon stations and tracker, transverse impact parameter with respect to the beam axis, and normalized $\chi^{2}$ of the global fit~\cite{CMS_wz}.
 The backgrounds from jets misidentified as muons and from semileptonic decays  of heavy quarks are suppressed by applying an isolation condition on the muon candidates. The isolation variable $I$ for muons is defined as
\ifthenelse{\boolean{cms@external}}
{
\begin{multline}
 I = \Bigl\{\Sigma \bigl[p_{T} (\text{tracks}) \\+ E_{T}(\mathrm{EM}) + E_{T}(\mathrm{HAD})\bigr] - \pi (\Delta R)^{2}\rho \Bigr\} /p_{T}^{\mu},
\end{multline}
}
{
\begin{equation}
 I = \left\{\Sigma \left[p_{T} (\text{tracks}) + E_{T}(\mathrm{EM}) + E_{T}(\mathrm{HAD})\right] - \pi (\Delta R)^{2}\rho \right\} /p_{T}^{\mu},
\end{equation}
}
where the sum is defined in a cone of radius \ifthenelse{\boolean{cms@external}}{\linebreak[4]}{}$\Delta R=\sqrt{(\Delta\phi)^{2} + (\Delta\eta)^{2}}=0.3$ around the muon direction; $\Delta\eta$ and $\Delta\phi$ are the pseudorapidity and azimuthal separation between the muon and tracks or calorimetric towers.
Here $p_{T}$(tracks) is the transverse momentum of tracks, excluding muons, with $p_{T}>$ 1\GeVc, $E_{T} ({\rm EM})$ is the transverse energy deposited in the electromagnetic calorimeter, $E_{T}({\rm HAD})$ is the transverse energy deposited in the hadronic calorimeter, and $\rho$ is the average energy density~\cite{salam:2008} in the calorimeter and tracker originating from additional inelastic pp interactions (pile-up) in the same bunch crossing as the DY interaction.The calculation of $\rho$ takes into account the number of reconstructed primary vertices in the event; the average value of $\rho$ is 5.6\GeVc.
 A muon is considered to be  isolated if $I < 0.15$. Because of the energy density correction, the isolation efficiency is independent of the number of pile-up interactions.

The selected muons are required to have opposite charges, transverse momenta larger than 20\GeVc, and pseudorapidity $|\eta| < 2.4$.
 Both muons are required to be associated with the same vertex, which is designated as the {\it signal vertex}.
 The selected signal vertex is required to be within $\pm$18\cm of the nominal interaction point as measured along the $z$ direction.
 At least five tracks are required to be associated with the signal vertex, and the transverse displacement of the signal vertex from the beam axis is required to be less than 2\cm.
 These criteria select a pure sample of DY events with a total background contribution of less than 0.5\% as estimated from simulated events.

Tracks, excluding the selected muons, are considered for the UE measurement if they are well reconstructed in the silicon-pixel and the silicon-strip tracker, have $p_{T} > 0.5\GeVc$ and $|\eta| < 2$, and originate from the signal vertex.
 To reduce the number of improperly reconstructed tracks, a {\it high purity} reconstruction algorithm~\cite{trk} is used. The high purity algorithm requires stringent cuts on the number of hits, the normalized $\chi^{2}$ of the track fit, and the consistency of the track originating from a pixel vertex.
 To reduce the contamination of secondary tracks from decays of long-lived particles and photon conversions, the distances of closest approach between the track and the signal vertex in the transverse plane and in the longitudinal direction are required to be less than 3 times the respective uncertainties.
 Tracks with poorly measured momenta are removed by requiring $\sigma(p_{T})/p_{T} < 5\%$, where $\sigma(p_{T})$ is the uncertainty on the $p_{T}$ measurement.
These selection criteria reject about 10\% of primary tracks and 95\% of misreconstructed and secondary tracks. The selected tracks have a contribution of about 2\% from misreconstructed and secondary tracks.

\subsection{Corrections and systematic uncertainties}
The UE observables, discussed in Section 2, are corrected for detector effects and selection efficiencies.
 The measured observables are corrected to reflect the activity from all primary charged particles with transverse momentum $p_{T} > 0.5\GeVc$ and pseudorapidity $|\eta| < 2$.
 The particle and energy densities are corrected using a {\it bin-by-bin} technique.
 In the bin-by-bin technique, the correction factor is calculated by taking the bin-by-bin ratio of the particle level and detector level distributions for simulated events and then the measured quantity is multiplied by this correction factor.
 There is a small growth in the particle and energy densities with increasing $p_{T}^{\mu\mu}$ and $M_{\mu\mu}$ in the towards and transverse regions.
 Because of this slow growth of densities the bin migration in $p_{T}^{\mu\mu}$ and $M_{\mu\mu}$ has a small effect on the measurements, therefore a bin-by-bin method is considered to be sufficiently precise.
 There is a fast rise in the energy and particle densities in the away region with the increase of  $p_{T}^{\mu\mu}$,  but corrected results using a bin-by-bin method are consistent with correction obtained from a Bayesian~\cite{unfold} technique.
 The transverse momenta of the charged particles  have very good resolution and are corrected using a bin-by-bin method.
 In this analysis the average of the calculated correction factors from \PYTHIA{}6 Z2, \PYTHIA{}6 D6T, and \MADGRAPH Z2 is used to correct the  experimental distributions.
The maximum deviation from the average correction factor is taken as the model-dependent systematic uncertainty, estimated to be 0.7--1.4\% for the particle and energy densities.
 In the case of charged-particle multiplicity, there is substantial bin migration and the corrected results using the Bayesian~\cite{unfold} and  bin-by-bin techniques differ by 10--15\%.
 Therefore the charged-particle multiplicity is corrected using a Bayesian unfolding technique with a response matrix obtained using the \PYTHIA{}6 Z2 tune.
 The systematic uncertainty related to the correction procedure is calculated by unfolding the data with response matrices obtained using different tunes.

In the analyzed data, there are on average 6--7 collisions in each bunch crossing.
Tracks originating from these pile-up interactions cause the UE activity to be overestimated,
so the measurements are corrected for the presence of pile-up interactions.
 The correction factor is calculated as the ratio of the UE activity for simulated events with and without pile-up.
 The uncertainty in the modelling of the pile-up events is estimated by varying the mean of the expected number of pile-up events by $\pm$1.
 This uncertainty in pile-up modelling affects the particle and energy densities by 0.3--1.0\%.
 The effect due to pile-up events is small because  only the tracks associated with the same vertex as the muon pair are used.
 The results are also cross-checked with low pile-up 7\TeV data collected during 2010 and the differences are found to be negligible.

\begin{table*}[htbp]
\caption{\label{tab:sys}{\small Summary of the systematic uncertainties on the particle and energy densities (in percent). The first three rows show the systematic uncertainties for the particle density in the towards, transverse, and away regions. The last three rows report the systematic uncertainties for the  energy density. The numbers outside the parentheses refer to the case where the densities are measured as a function of $M_{\mu\mu}$ and those in the parentheses correspond to the measurements as a function of  $p_{T}^{\mu\mu}$.} }
\begin{center}
{\footnotesize
 \begin{tabular}{l|c|c|c|c|c|c}\hline
 Observable  & model & pile-up & isolation & mis-ID & background & total\\
             &  $M_{\mu\mu}$ ($p_{T}^{\mu\mu}$) &  $M_{\mu\mu}$ ($p_{T}^{\mu\mu}$) &  $M_{\mu\mu}$ ($p_{T}^{\mu\mu}$) &  $M_{\mu\mu}$ ($p_{T}^{\mu\mu}$) &  $M_{\mu\mu}$ ($p_{T}^{\mu\mu}$) &  $M_{\mu\mu}$ ($p_{T}^{\mu\mu}$)\\
\hline
$1/[\Delta\eta \Delta(\Delta\phi)] \langle N_{\rm ch} \rangle$ (towards)  & 0.8 (0.8) & 1.0 (0.9) & 0.9--1.5 (0.9) & 1.0 (1.0) & 0.7 (0.3) & 2.0--2.3 (1.8)\\
$1/[\Delta\eta \Delta(\Delta\phi)] \langle N_{\rm ch} \rangle$ (transverse) & 0.7 (0.9) & 0.9 (0.9) & 0.8--1.7 (0.8) & 0.9 (0.9) & 0.7 (0.5) & 1.8--2.3 (1.8)\\
$1/[\Delta\eta \Delta(\Delta\phi)] \langle N_{\rm ch} \rangle$ (away)  & 0.7 (0.6) & 0.9 (0.3--0.9) & 0.8--1.6 (0.8) & 0.9 (0.9) & 0.5 (0.5) & 1.7--2.2 (1.5--1.7)\\
\hline
$1/[\Delta\eta \Delta(\Delta\phi)] \langle \Sigma p_{T} \rangle$ (towards)  & 1.2 (1.2) & 0.8 (0.7) & 1.1--2.0 (1.4) & 0.8 (0.8) & 0.8 (0.7) & 2.1--2.7 (2.2)\\
$1/[\Delta\eta \Delta(\Delta\phi)] \langle \Sigma p_{T} \rangle$ (transverse) & 1.1 (1.4) & 0.7 (0.7) & 1.0--2.5 (1.3) & 0.8 (0.8) & 0.8 (0.9) & 2.0--3.0 (2.4)\\
$1/[\Delta\eta \Delta(\Delta\phi)] \langle \Sigma p_{T} \rangle$ (away)  & 1.0 (0.8) & 0.7 (0.3--0.7) &  1.1--2.2 (1.1) & 0.8 (0.7) & 0.7 (0.2) & 2.0--2.7 (1.6--1.7)\\
\hline
\end{tabular}
}
\end{center}
\end{table*}

We also consider possible systematic effects related to trigger requirements, different beam-axis positions in data and simulation, various track selection criteria, muon isolation, and misidentification of tracks.
The combined systematic uncertainty related to trigger conditions, the varying beam-axis position, and track selection is less than 0.5\%.
 The systematic uncertainty due to isolation is calculated by removing the isolation condition in the simulated events used for the correction and is found to be 0.8--2.5\% for the particle and energy densities.

The yield of secondary tracks originating from the decay of long-lived particles is not correctly predicted by the simulation~\cite{strange}.
 To estimate the effect of secondary tracks, a subset of simulated events is created by rejecting tracks that do not have a matching primary charged particle at the generator level.
 The uncertainty is evaluated by correcting the measurements with this subset of the simulated events, containing fewer secondary tracks, and is found to be 0.7--1.0\% for the particle and energy densities.

Though the total contribution of background processes is very small, it affects the measurement at higher $p_{T}^{\mu\mu}$ (50--100\GeVc) and small $M_{\mu\mu}$ (40--60\GeVcc) where the contamination from  \cPqt\cPaqt\ and DY$\rightarrow\tau\tau$ background processes is 1\% and 5\%, respectively.
 The particle and energy densities differ between DY$\rightarrow\tau\tau$ and DY$\rightarrow\mu\mu$ (the signal process) by 20\%.
 The particle (energy) density for the \cPqt\cPaqt\ background is two times (four times) that for the signal process.
 Combination of the differences in the densities for background processes and relative background contributions gives a systematic uncertainty of 0.2--0.9\%.

Table~\ref{tab:sys} summarizes the dominant systematic uncertainties on the particle and energy densities.
 The total systematic uncertainty on the particle and energy densities is in the range 1.5--3.0\%, whereas the uncertainties on the track multiplicity and $p_{T}$ spectra reach 10\% in the tail (not reported in Table~\ref{tab:sys}).
In all figures, inner error bars represent the statistical uncertainty only, while outer error bars account for the quadratic sum of statistical and systematic uncertainties.

\section{Results}
The UE activity in DY events, for charged particles with $p_{T} > 0.5\GeVc$ and $|\eta| < 2.0$, is presented as a function of $M_{\mu\mu}$ and $p_{T}^{\mu\mu}$.
 The multiplicity and the transverse momentum distributions are also presented for two different sets of events, $p_T^{\mu\mu} < 5\GeVc$ and  $81 < M_{\mu\mu} < 101\GeVcc$.
 Finally, the UE activity in the transverse region is compared with that measured in hadronic events using a leading track-jet.
\subsection{Underlying event in the Drell--Yan process}

\begin{figure*}[htbp]
\begin{center}
\begin{tabular}{@{}c@{}@{}c@{}@{}c@{}}
 \multicolumn{1}{c}{\scriptsize particle density} & \multicolumn{1}{c}{\scriptsize energy density} &  \multicolumn{1}{c}{\scriptsize ratio of energy and particle densities}\\
  \includegraphics[width=0.32\textwidth]{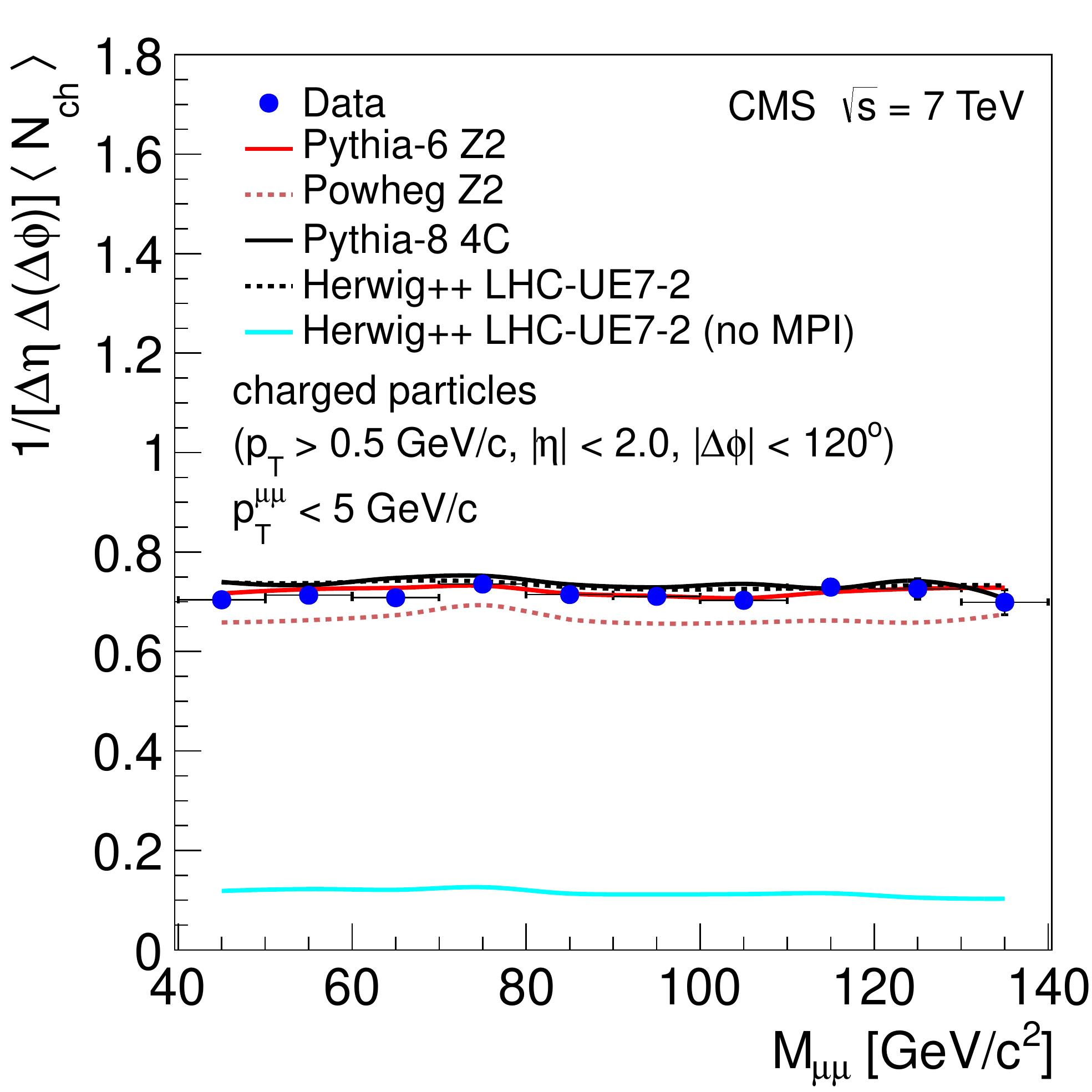} & \includegraphics[width=0.32\textwidth]{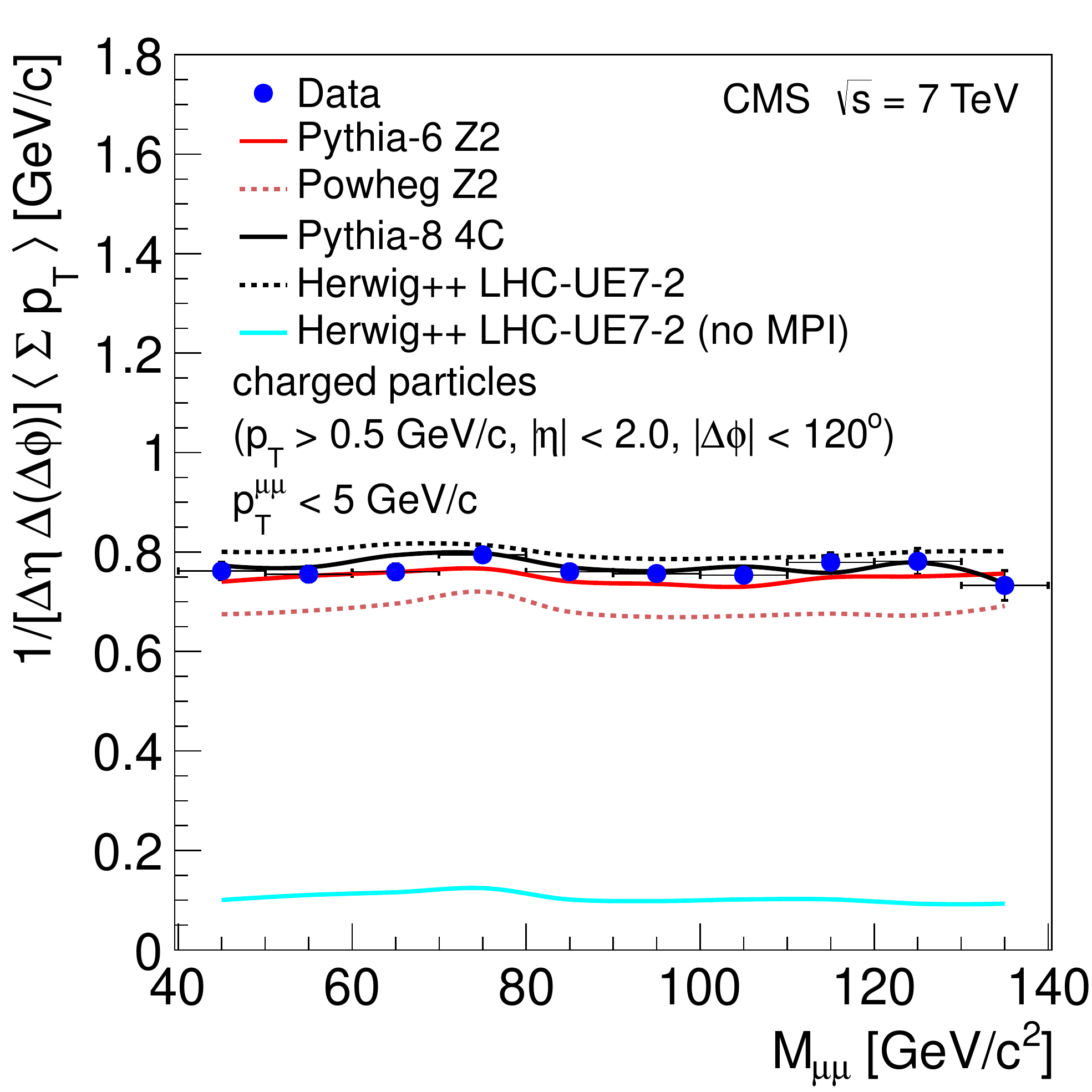} & \includegraphics[width=0.32\textwidth]{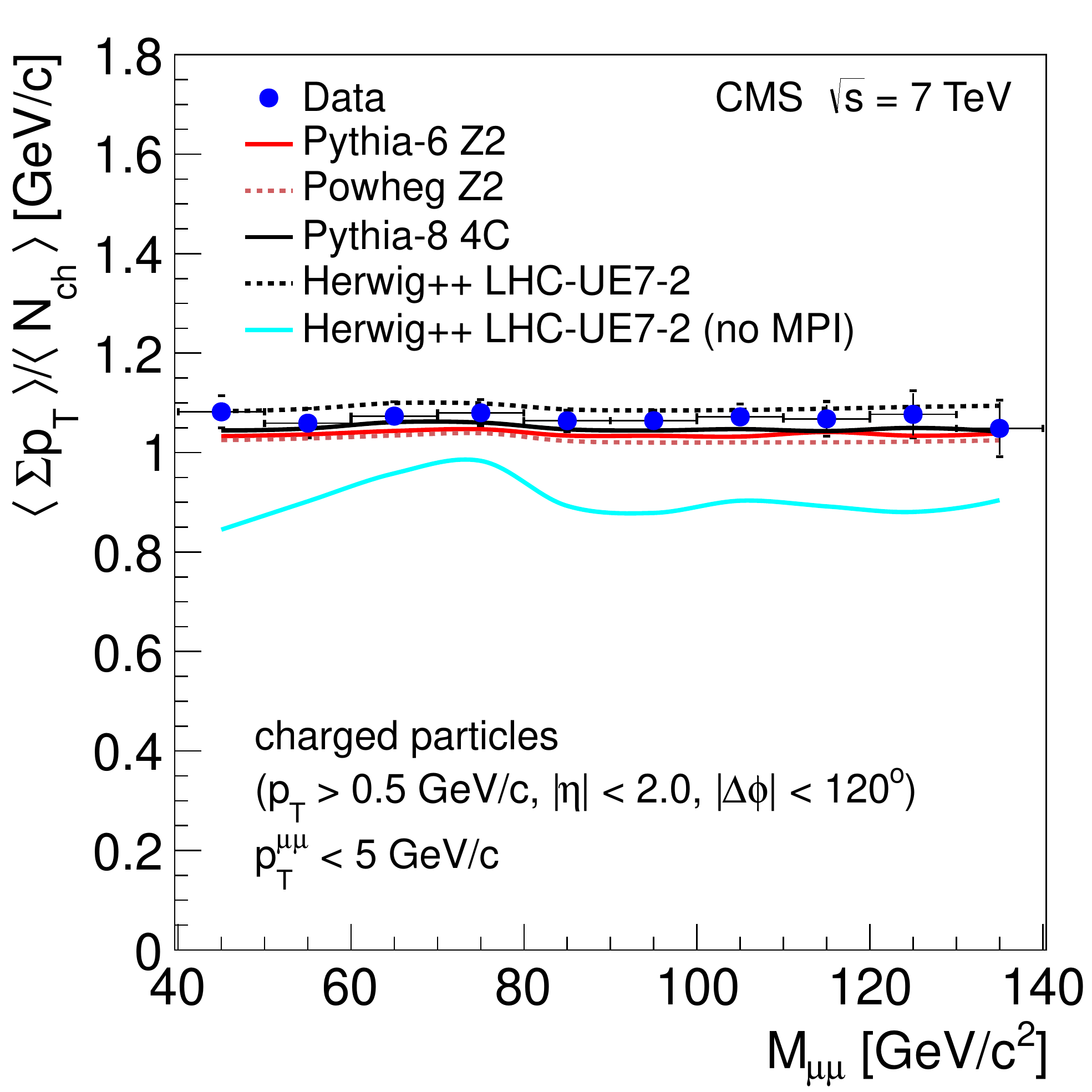}\\
  \includegraphics[width=0.32\textwidth]{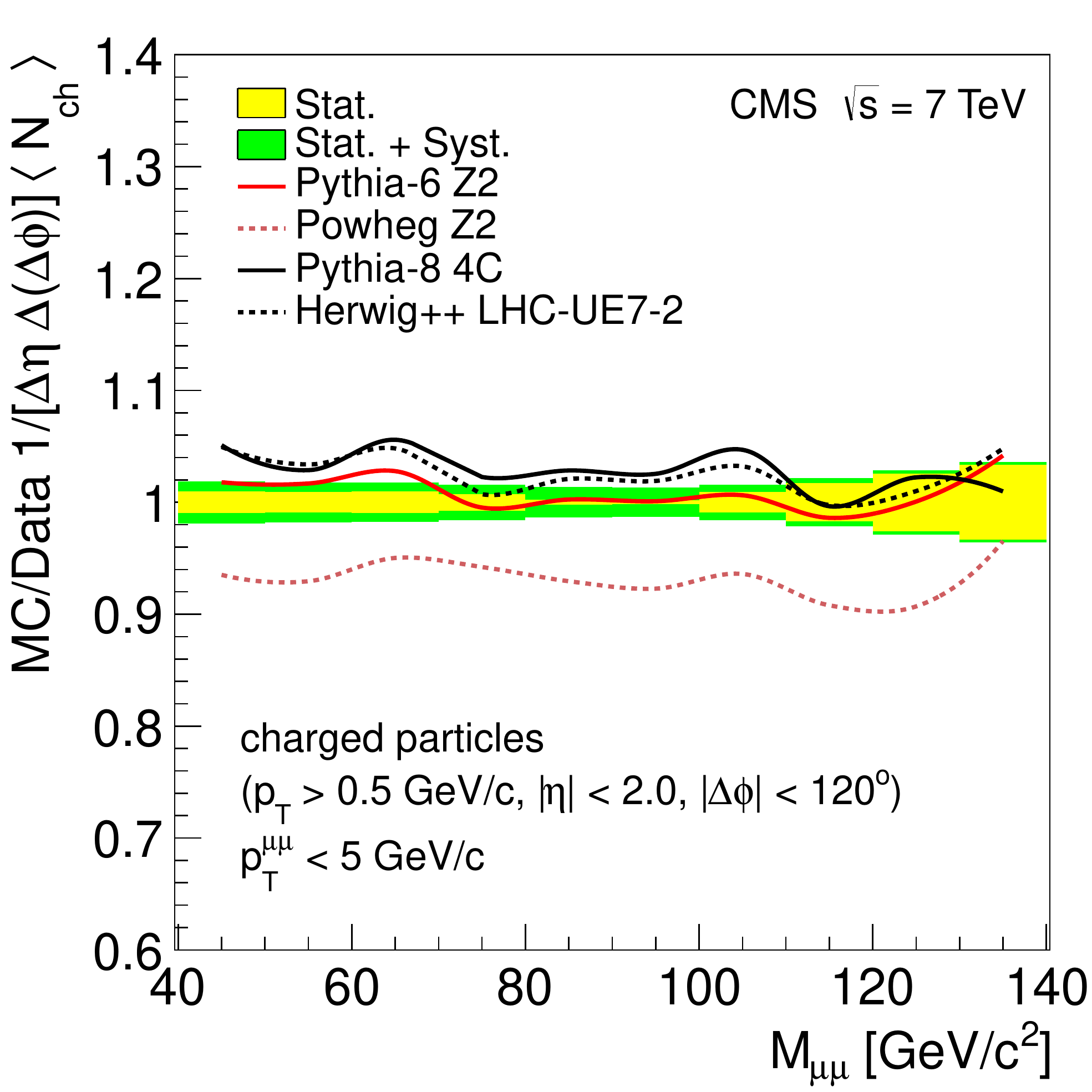} & \includegraphics[width=0.32\textwidth]{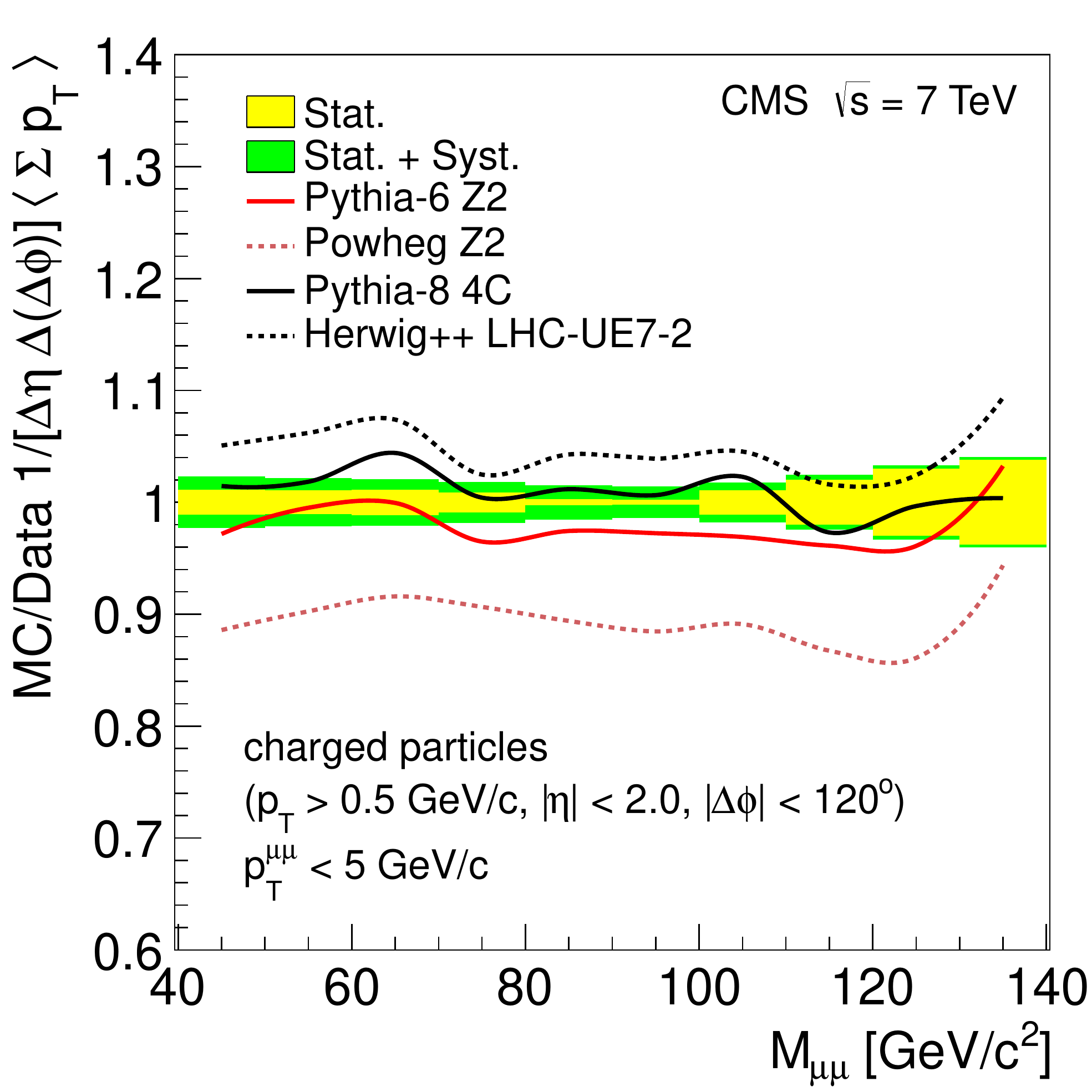} & \includegraphics[width=0.32\textwidth]{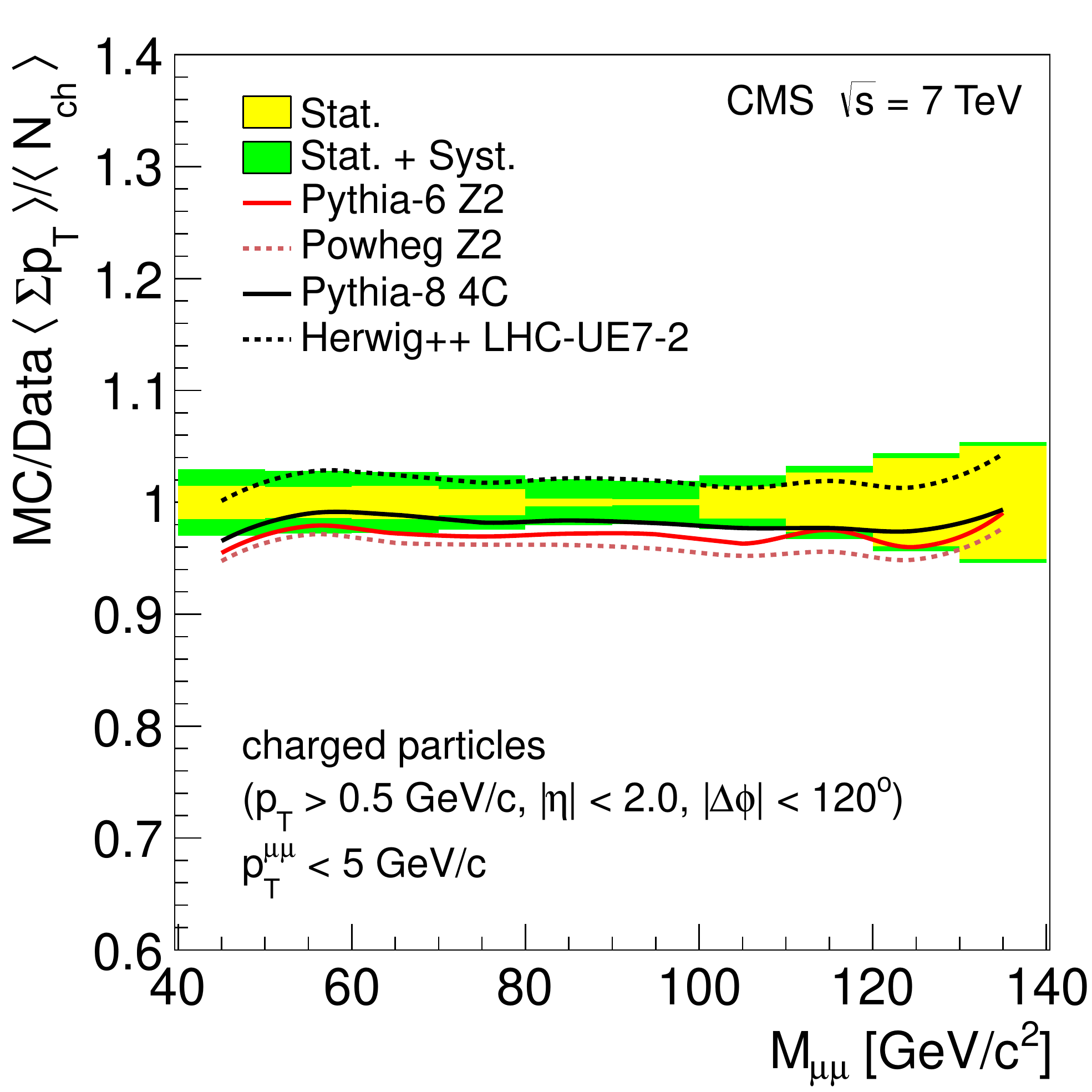}\\
 \end{tabular}
\end{center}
  \caption{Top: The UE activity as a function of the dimuon invariant mass ($M_{\mu\mu}$) for events with $p_{T}^{\mu\mu} < 5$\GeVc for charged particles having $\Delta\phi < 120^{\circ}$: (left) particle density; (centre) energy density; (right) ratio of the energy and particle densities. The predictions of \PYTHIA{}6 Z2, \POWHEG Z2, \PYTHIA{}8 4C, and \HERWIG{}++ LHC-UE7-2 (with and without MPIs) are also displayed. In the top right plot, the structure around 60--80\GeVcc for \HERWIG{}++ without MPIs reflects the influence of photon radiation by final-state muons, which is enhanced below the \cPZ\ resonance. Bottom: Ratios of the predictions of various MC models and the measurement. The inner band shows the statistical uncertainity of data whereas the outer band represents the total uncertainty.}
\label{fig:profile_mll}
\end{figure*}

The energy-scale dependence of the MPI activity is studied by limiting the ISR.
 To accomplish this we require the muons to be back-to-back in the transverse plane with $p_T^{\mu\mu} < 5\GeVc$ and measure the dependence of the UE activity on the dimuon mass, $M_{\mu\mu}$.
 The resulting particle and energy densities are shown in Fig.~\ref{fig:profile_mll}.
 Because the activity is almost identical in the towards and transverse regions, they are combined as $|\Delta\phi| < 120^{\circ}$.
 The contribution of ISR to the UE activity is small after requiring $p_{T}^{\mu\mu} < 5\GeVc$, as shown by the prediction of \HERWIG{}++ without MPIs.
 This figure also illustrates the dominant role of MPIs in our current models as they generate more than 80\% of the UE activity in these ISR-reduced events.
 The lack of  dependence of the UE activity on $M_{\mu\mu}$ within the range under study (40--140\GeVcc) indicates that the activity due to MPIs is constant at energy scales down to 40\GeV.
 The quantitative description by model tunes based on the minimum-bias and UE observables in hadronic events is illustrated by the MC/Data ratios in Fig.~\ref{fig:profile_mll}.
 In general, \PYTHIA{}6 Z2, \PYTHIA{}8 4C, and \HERWIG{}++ LHC-UE7-2 describe the densities well, whereas the Z2 tune used together with the \POWHEG generator underestimates both densities by 5--15\%.
 Both \PYTHIA and \HERWIG{}++ model tunes derived from the UE measurement in hadronic events using the leading jet/track approach describe the UE activity in the Drell--Yan events equally well and hence illustrate a certain universality of the underlying event across QCD and electroweak processes in hadronic collisions.

\begin{figure*}[htbp]
\centering
\begin{tabular}{@{}c@{}@{}c@{}@{}c@{}}
 \multicolumn{1}{c}{\scriptsize particle density} & \multicolumn{1}{c}{\scriptsize energy density} &  \multicolumn{1}{c}{\scriptsize ratio of energy and particle densities}\\
 \includegraphics[width=0.325\textwidth]{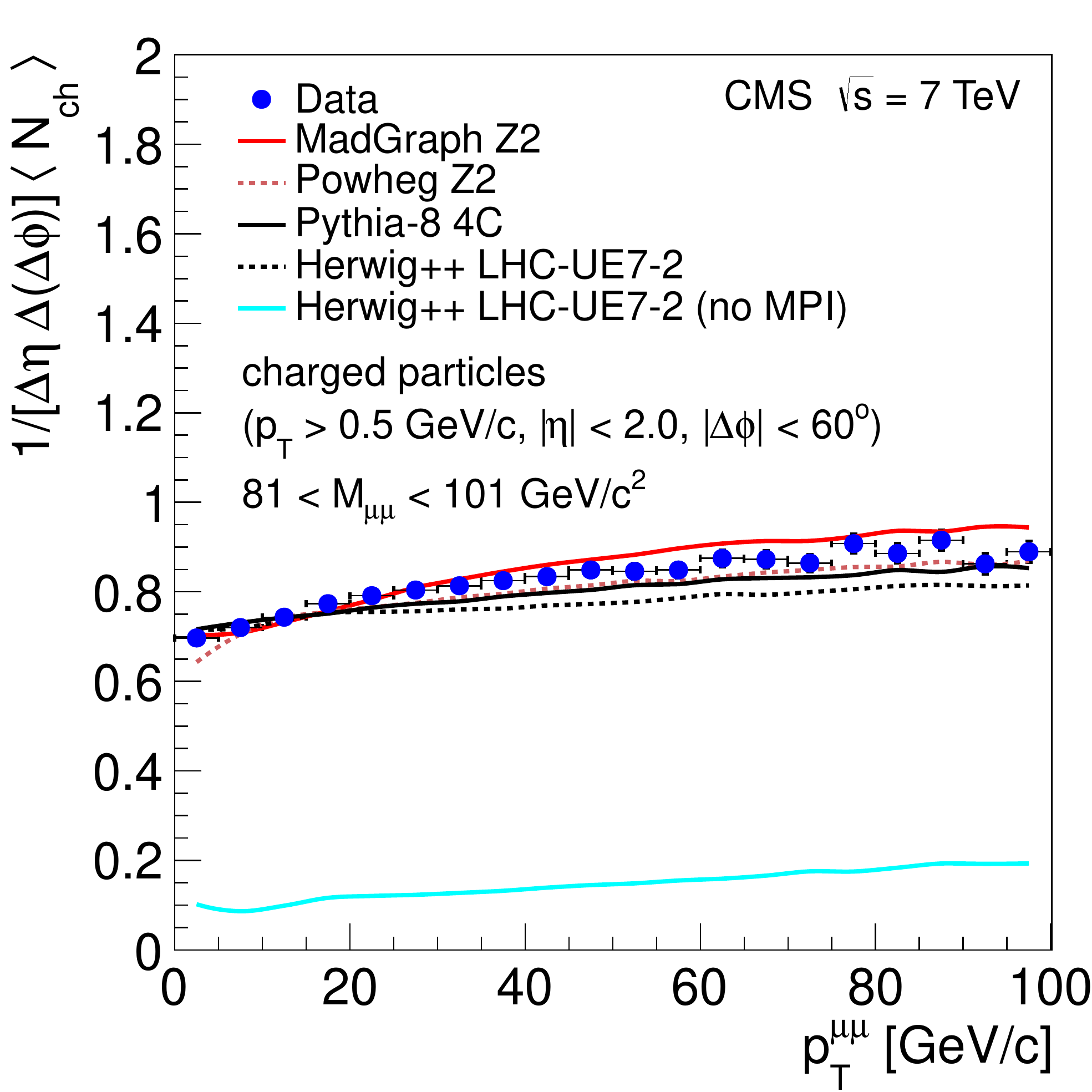} & \includegraphics[width=0.325\textwidth]{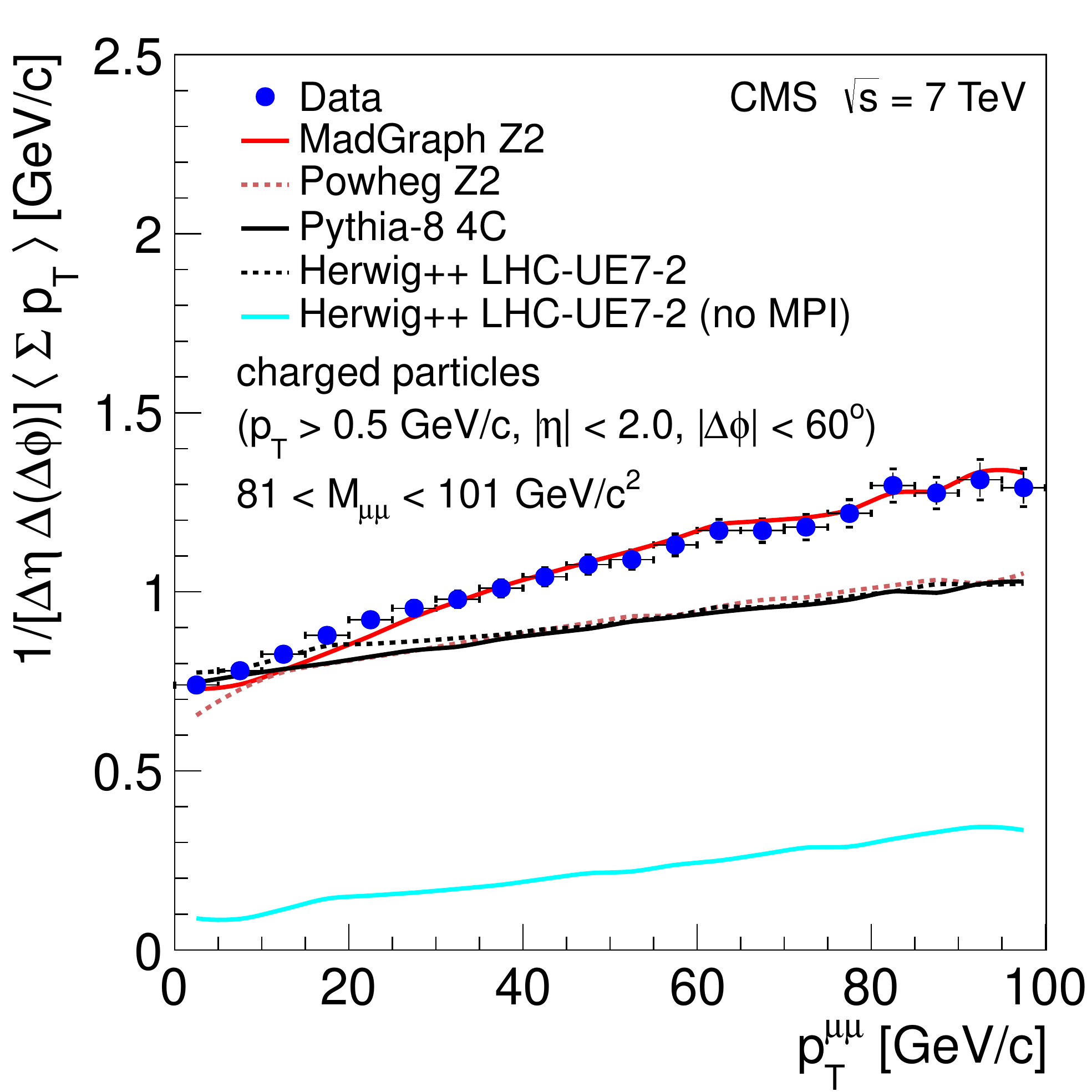} & \includegraphics[width=0.325\textwidth]{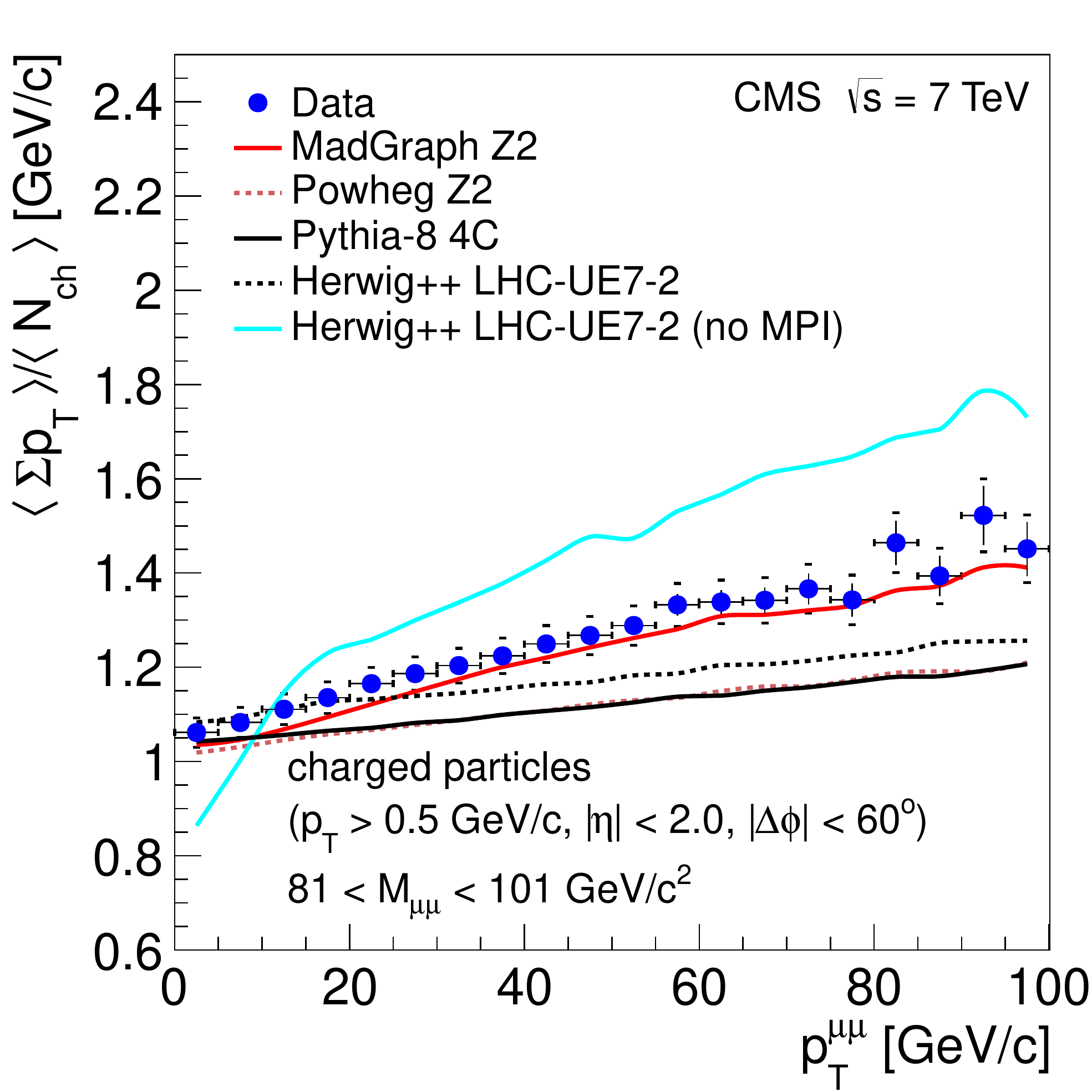}\\
 \includegraphics[width=0.325\textwidth]{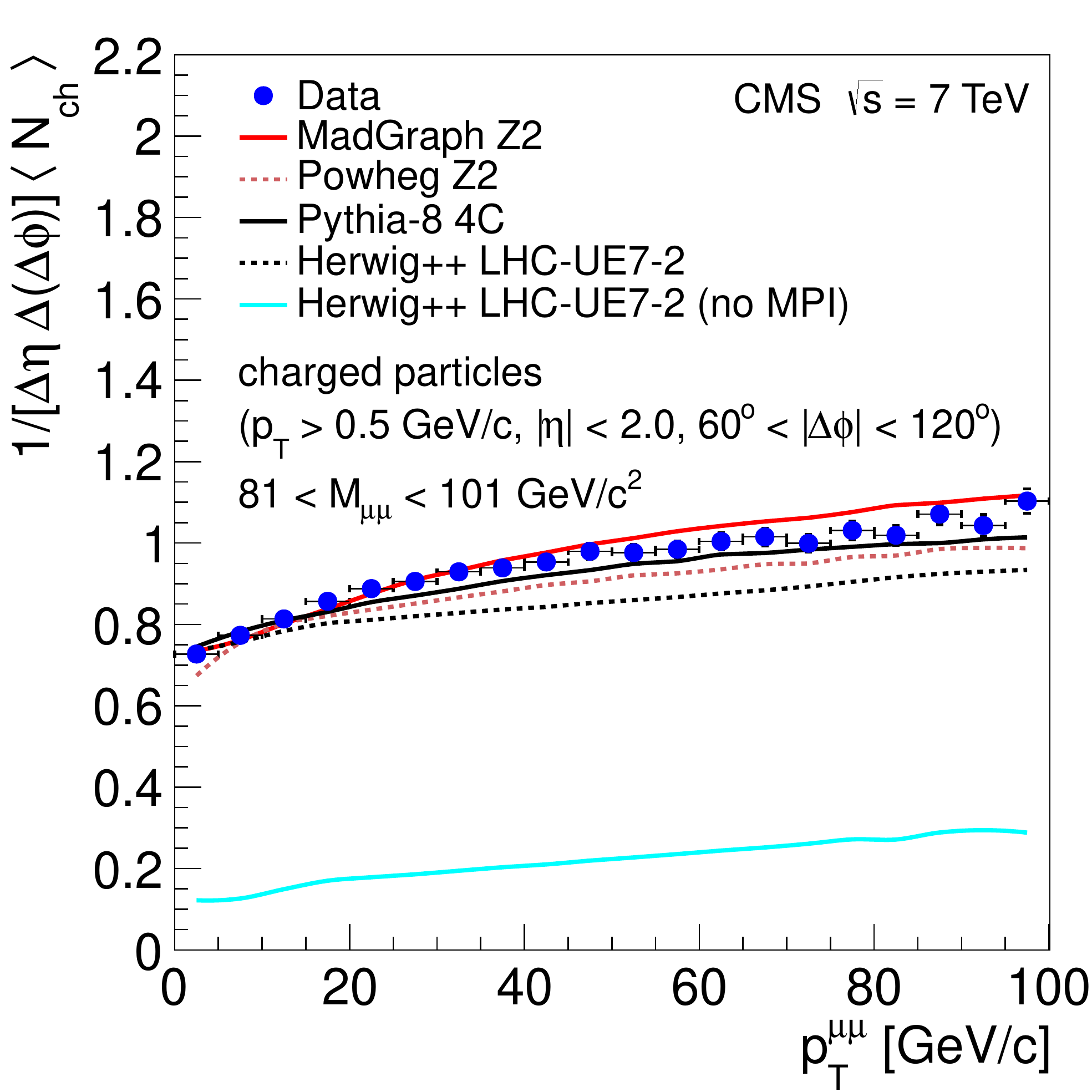} & \includegraphics[width=0.325\textwidth]{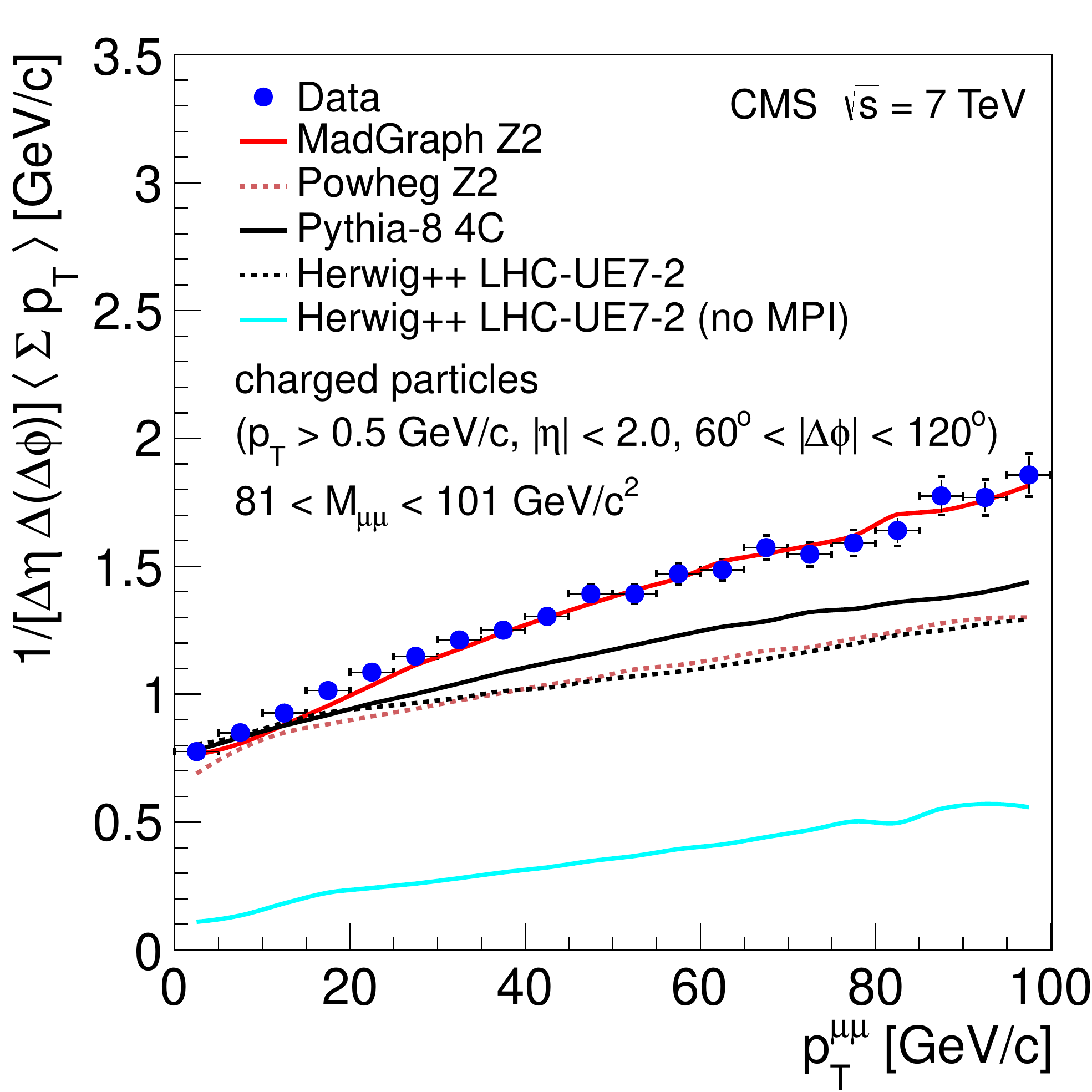} & \includegraphics[width=0.325\textwidth]{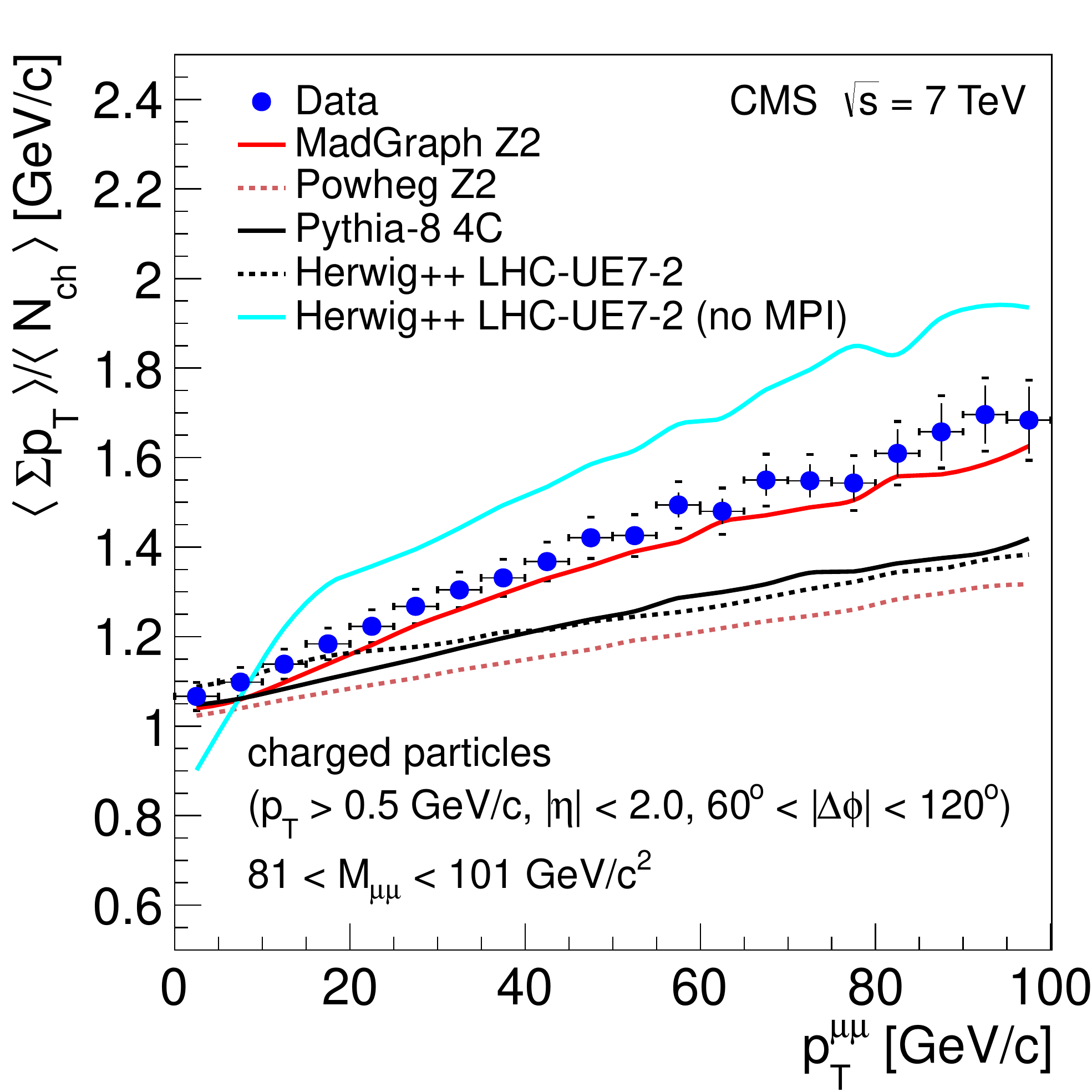}\\
 \includegraphics[width=0.325\textwidth]{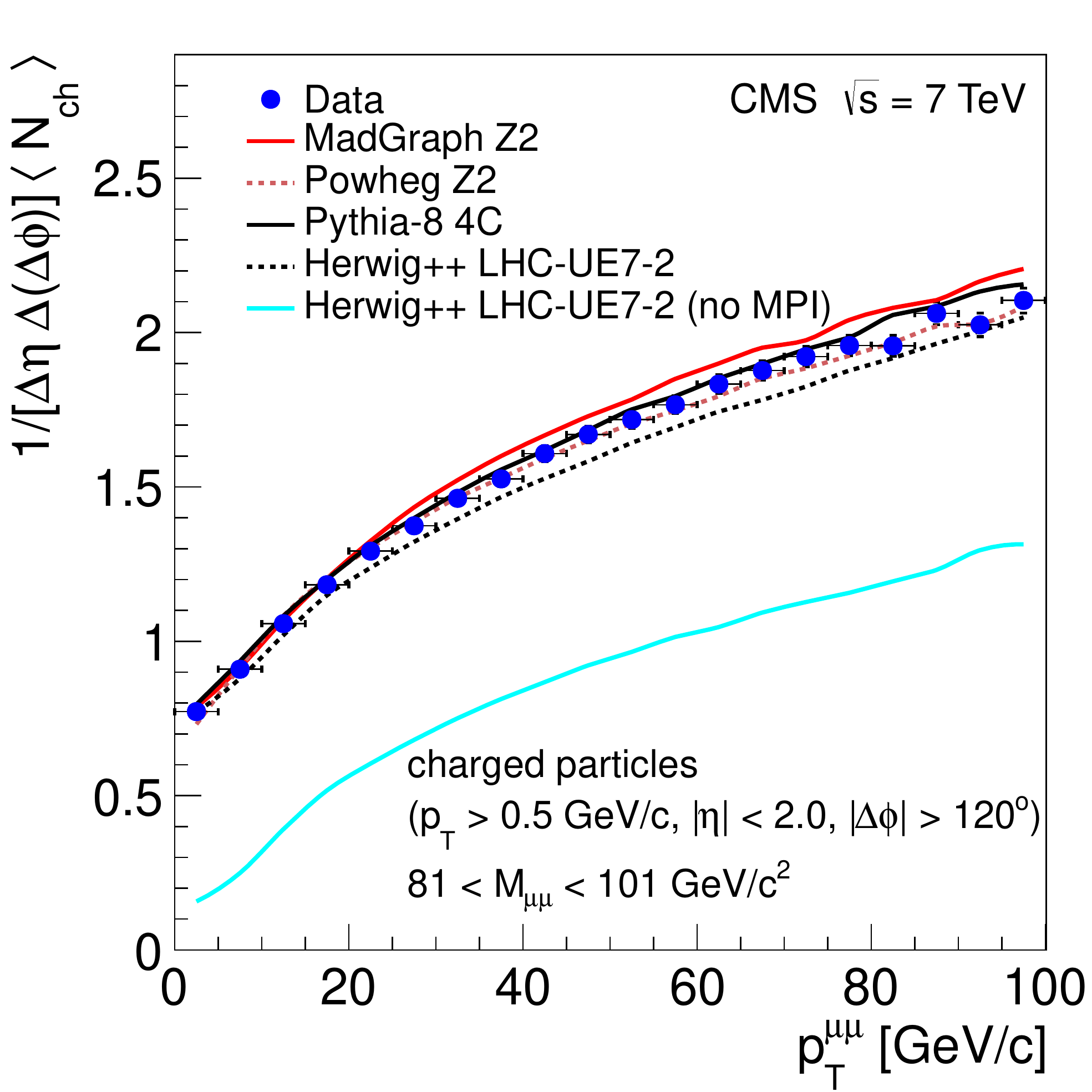} & \includegraphics[width=0.325\textwidth]{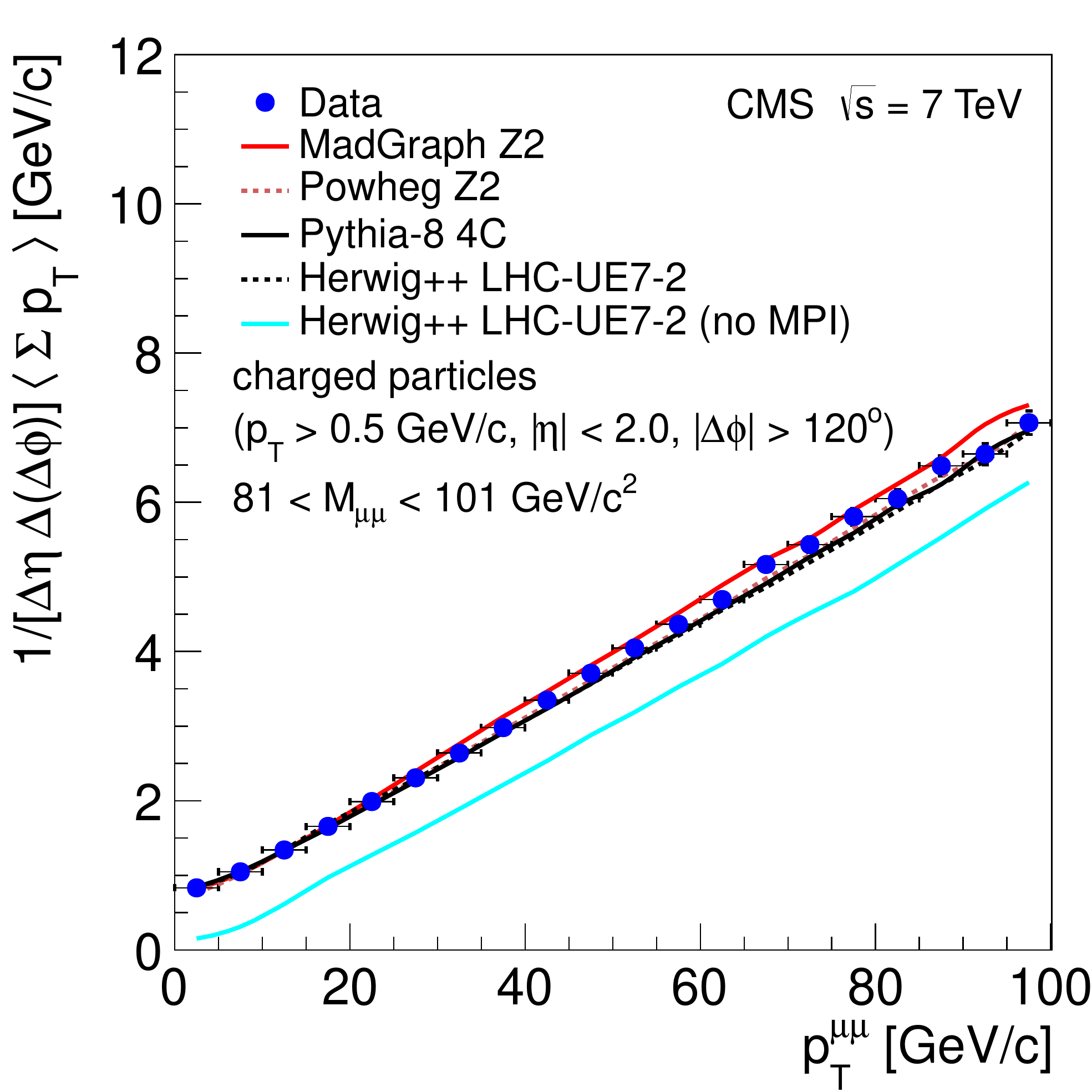} & \includegraphics[width=0.325\textwidth]{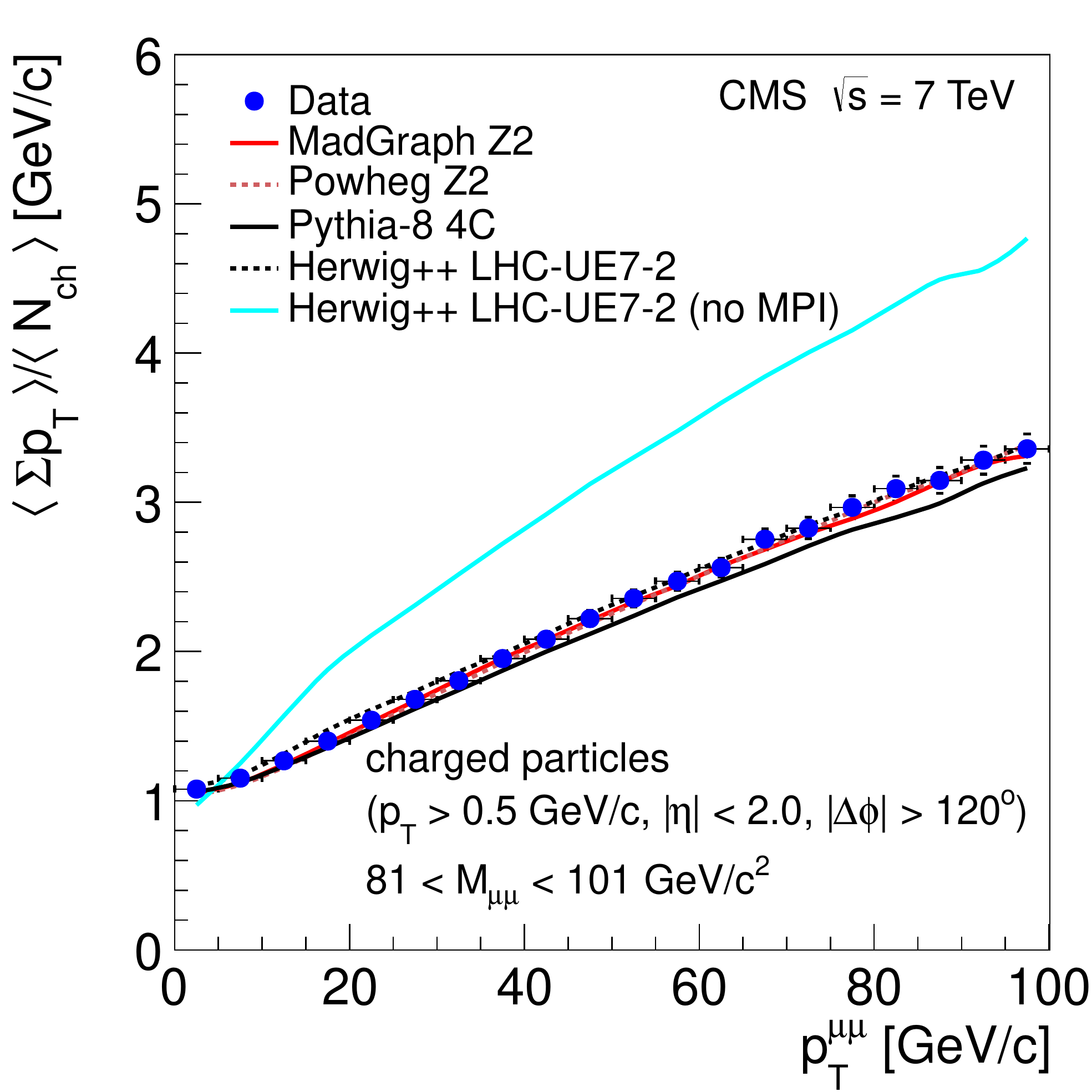}\\
 \end{tabular}
 \caption{The UE activity in the towards (upper row),  transverse (centre row), and away (bottom row) regions as functions of $p_{T}^{\mu\mu}$ for events satisfying $81 < M_{\mu\mu} < 101$\GeVcc: (left) particle density;
 (centre) energy density; (right) the ratio of the energy density and the particle densities. Predictions of \MADGRAPH Z2, \POWHEG Z2, \PYTHIA{}8 4C, and \HERWIG{}++ LHC-UE7-2 (with and without MPIs) are superimposed.}
\label{fig:profile_pt}
\end{figure*}

Dependence of the UE activity on the transverse momentum of the dimuon system is shown in Fig.~\ref{fig:profile_pt} in the towards, transverse, and away regions (top to bottom) for events having  $M_{\mu\mu}$ between 81\GeVcc and 101\GeVcc.
 At this high energy scale, the $p_{T}^{\mu\mu}$ dependence of the UE activity is sensitive to the ISR.
 The slope in the $p_{T}^{\mu\mu}$ dependence of the UE activity is identical for a model with and without MPIs and is therefore mainly due to ISR.
 The predictions of \HERWIG{}++ without MPIs underestimate the measurements in the away region as well because the MPIs produce particles uniformly in all directions.
 The UE activity does not fall to zero when $p_{T}^{\mu\mu}\rightarrow0$ because of the presence of the hard scale set by $M_{\mu\mu}$.

The particle and energy densities in the away region rise sharply with $p_{T}^{\mu\mu}$ and, because of momentum conservation mainly sensitive to the spectrum of the hardest emission, are equally well described by all tunes and generators considered.
 In the towards and transverse regions there is a slow growth in the particle and energy densities with increasing $p_{T}^{\mu\mu}$.
 The energy density increases more than the particle density, implying a continuous increase in the average transverse momentum of the charged particles with $p_{T}^{\mu\mu}$.
 This effect is also reflected in the ratio of the energy density to the particle density.
  The activity in the towards region is qualitatively similar to that in the transverse region.
 Quantitatively, the activity is higher in the transverse region than the towards region, an effect caused by the spill-over contributions from the recoil activity in the away region, which balances the dimuon system.
 This observation is visible in Fig.~\ref{fig:profile_pt} at small $p_{T}^{\mu\mu}$, where the radiation contribution is small and the activity in the transverse region is the same as that in the towards region.

\begin{figure*}[htbp]
\centering
\begin{tabular}{@{}c@{}@{}c@{}@{}c@{}}
\multicolumn{1}{c}{\scriptsize particle density} & \multicolumn{1}{c}{\scriptsize energy density} &  \multicolumn{1}{c}{\scriptsize ratio of energy and particle densities}\\
\includegraphics[width=0.325\textwidth]{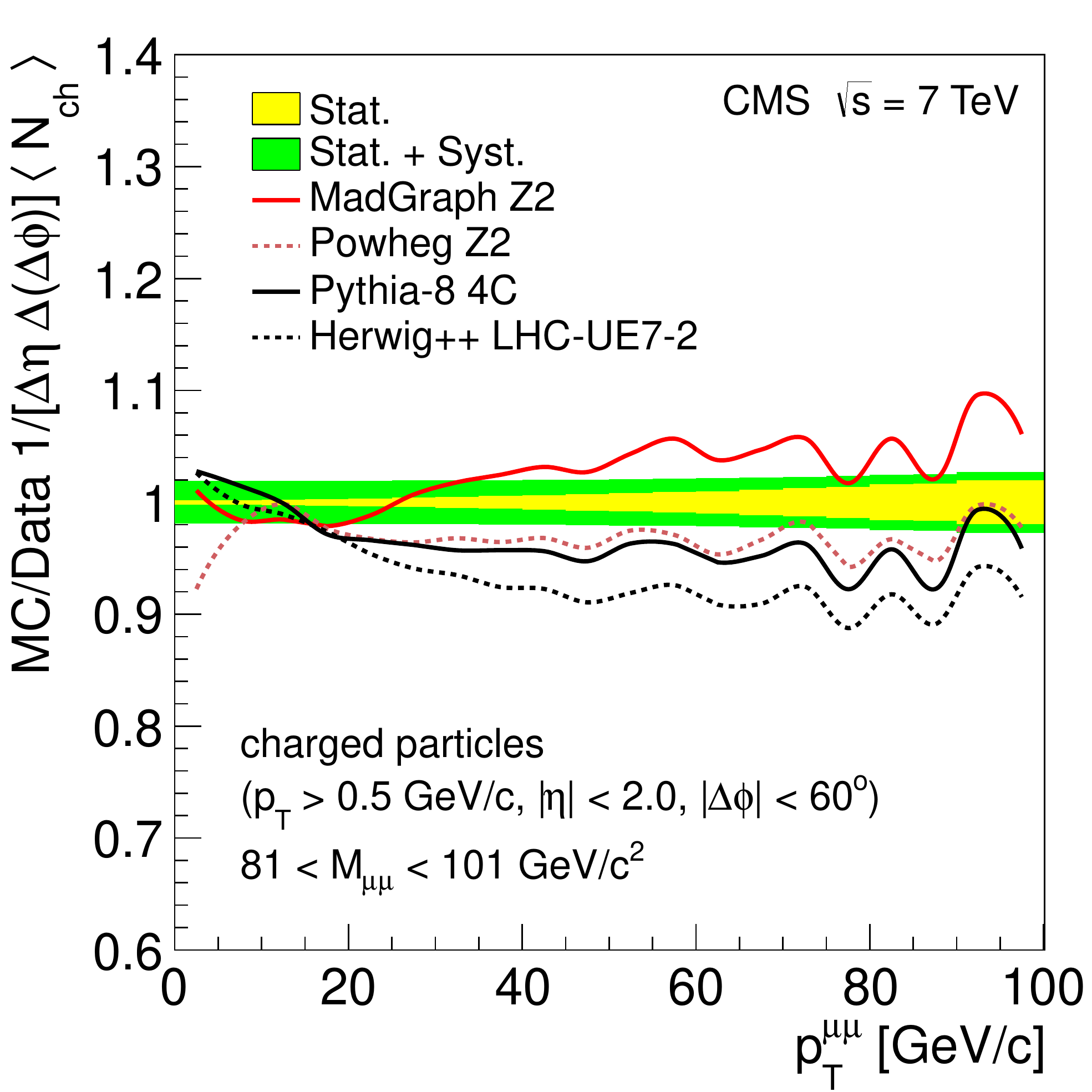} & \includegraphics[width=0.325\textwidth]{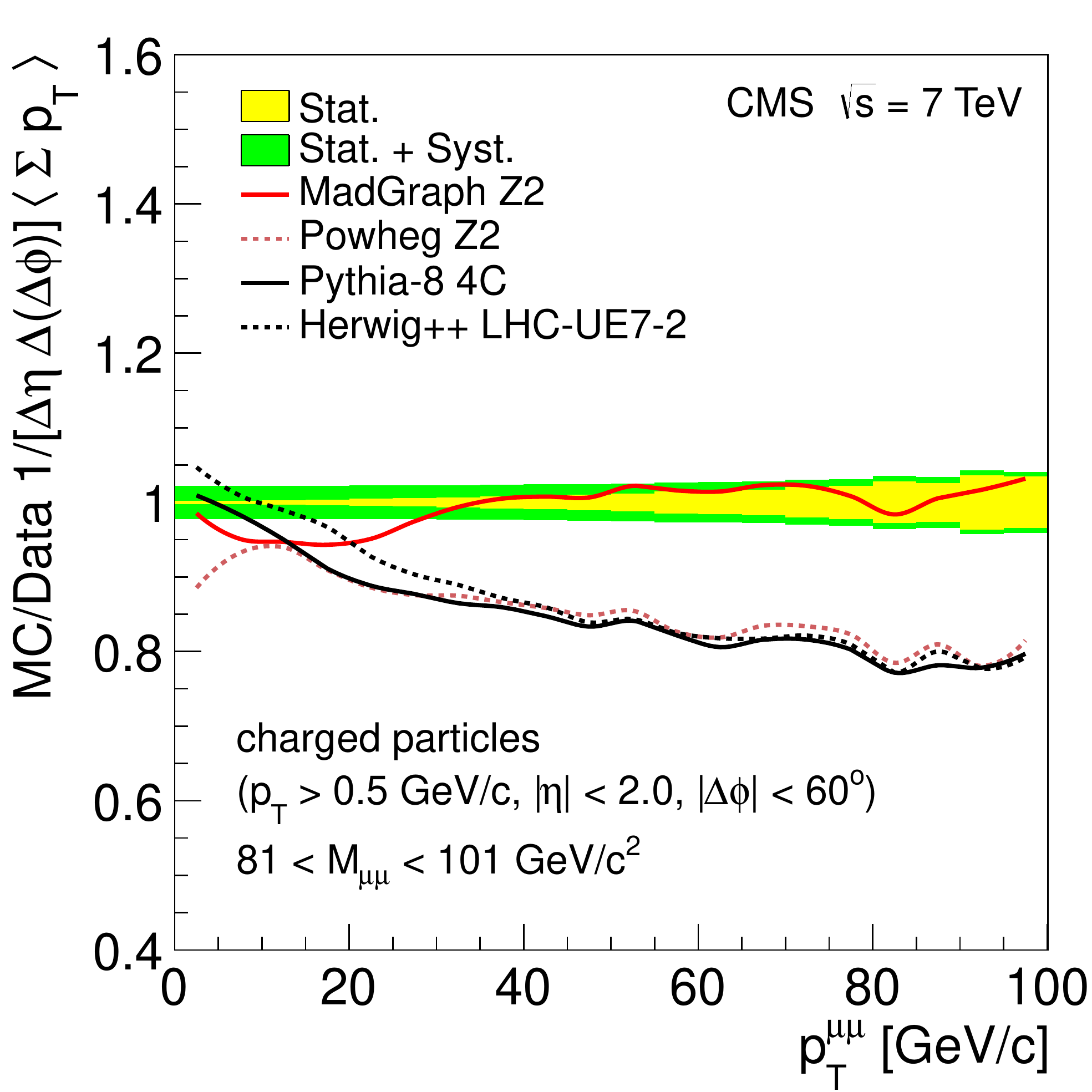} & \includegraphics[width=0.325\textwidth]{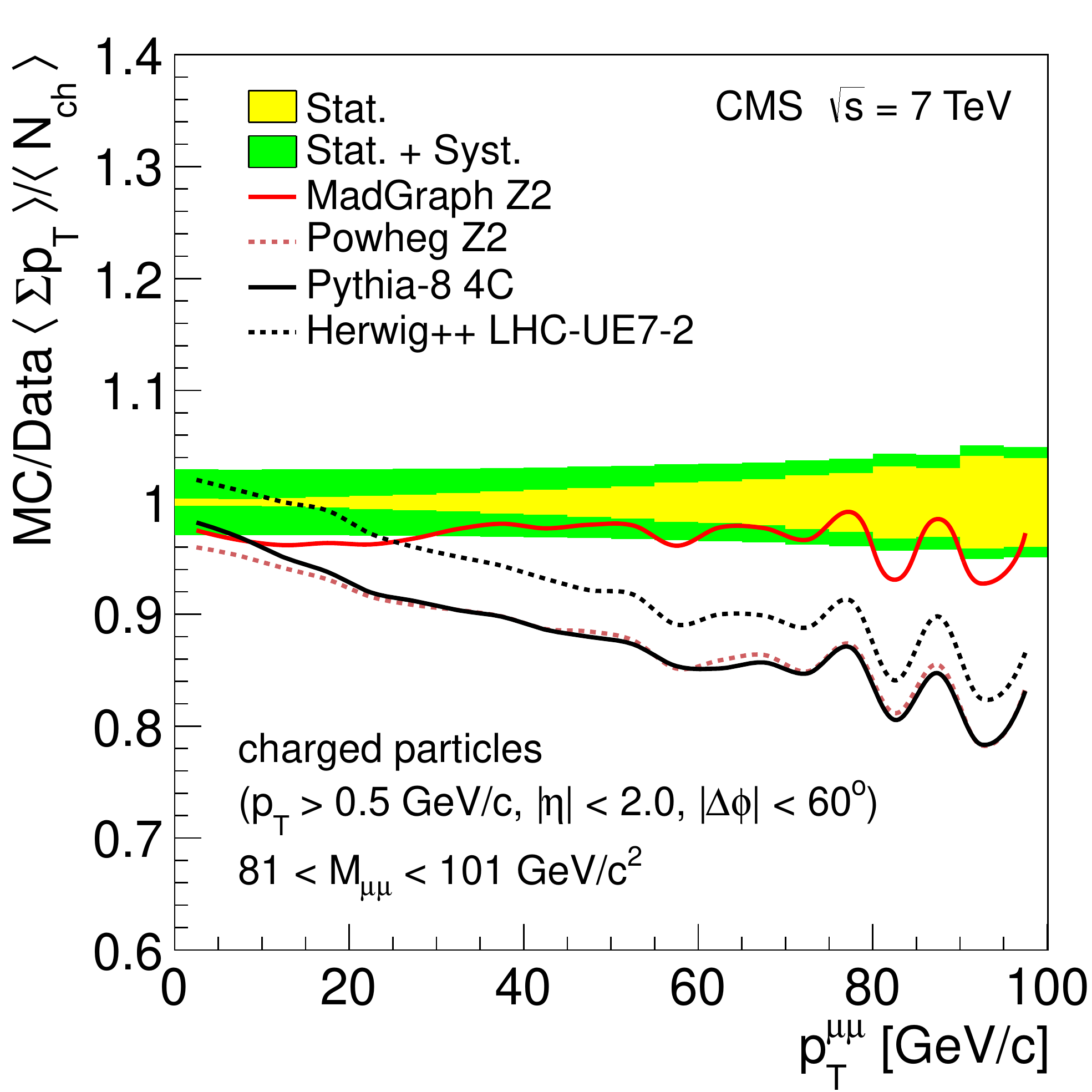}\\
\includegraphics[width=0.325\textwidth]{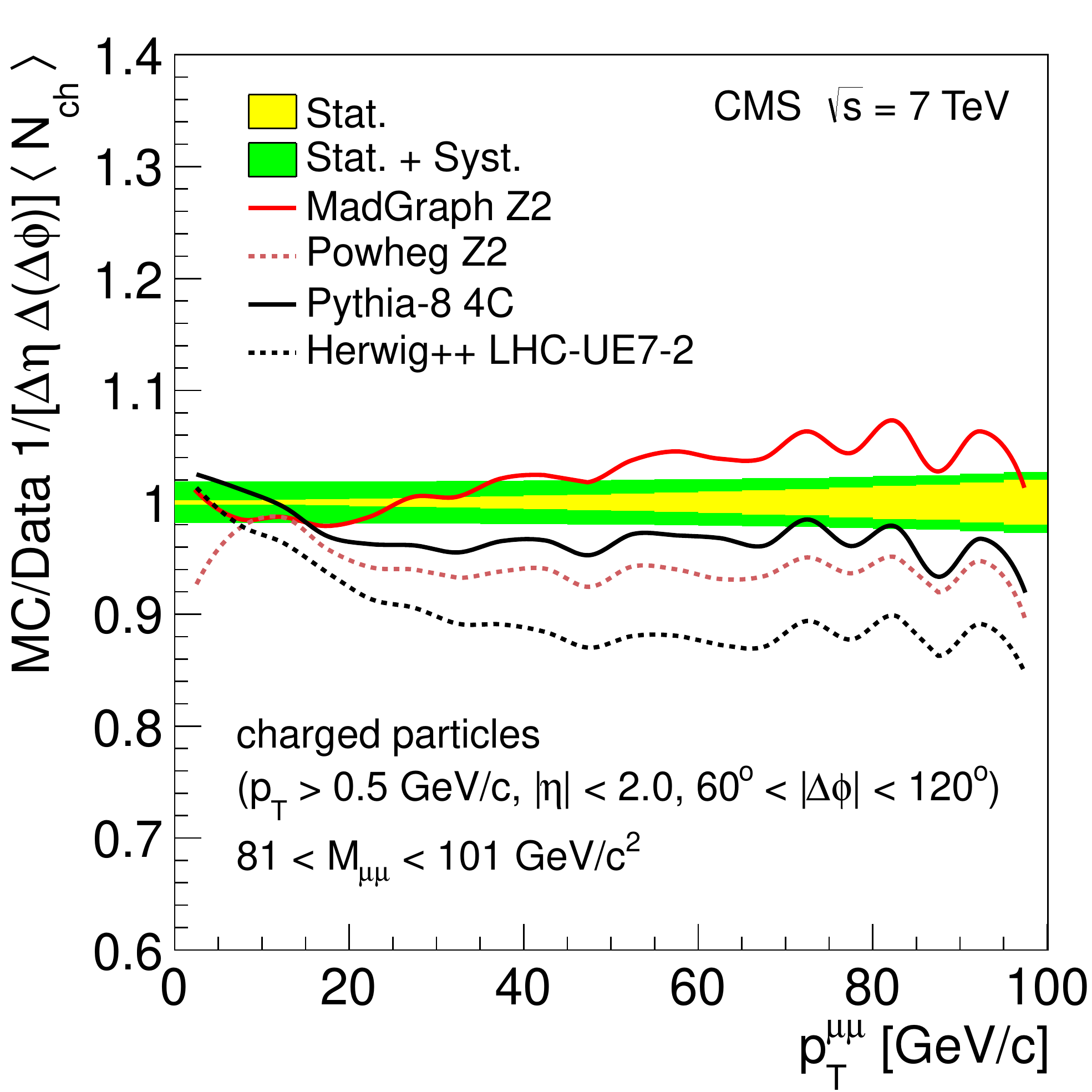} & \includegraphics[width=0.325\textwidth]{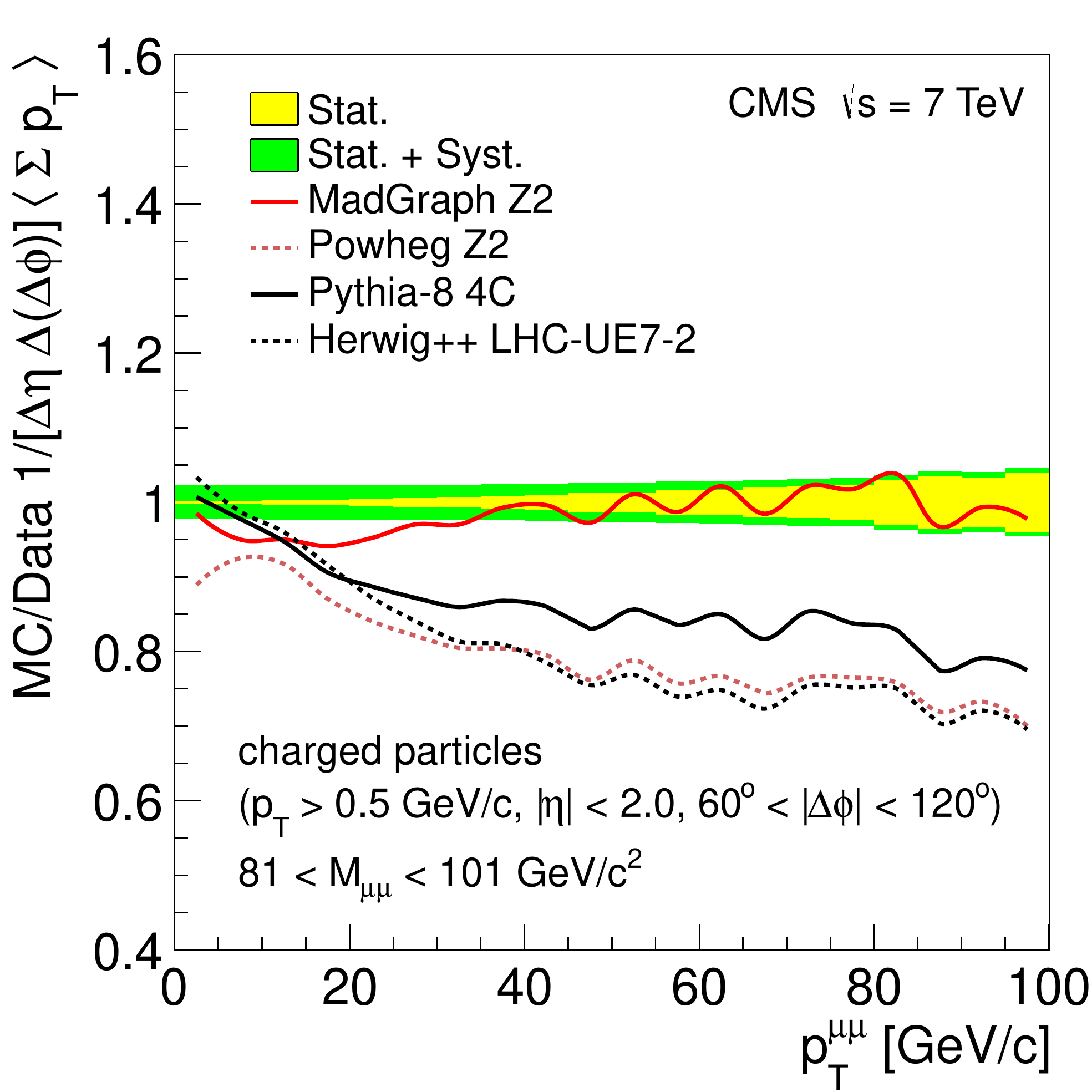} & \includegraphics[width=0.325\textwidth]{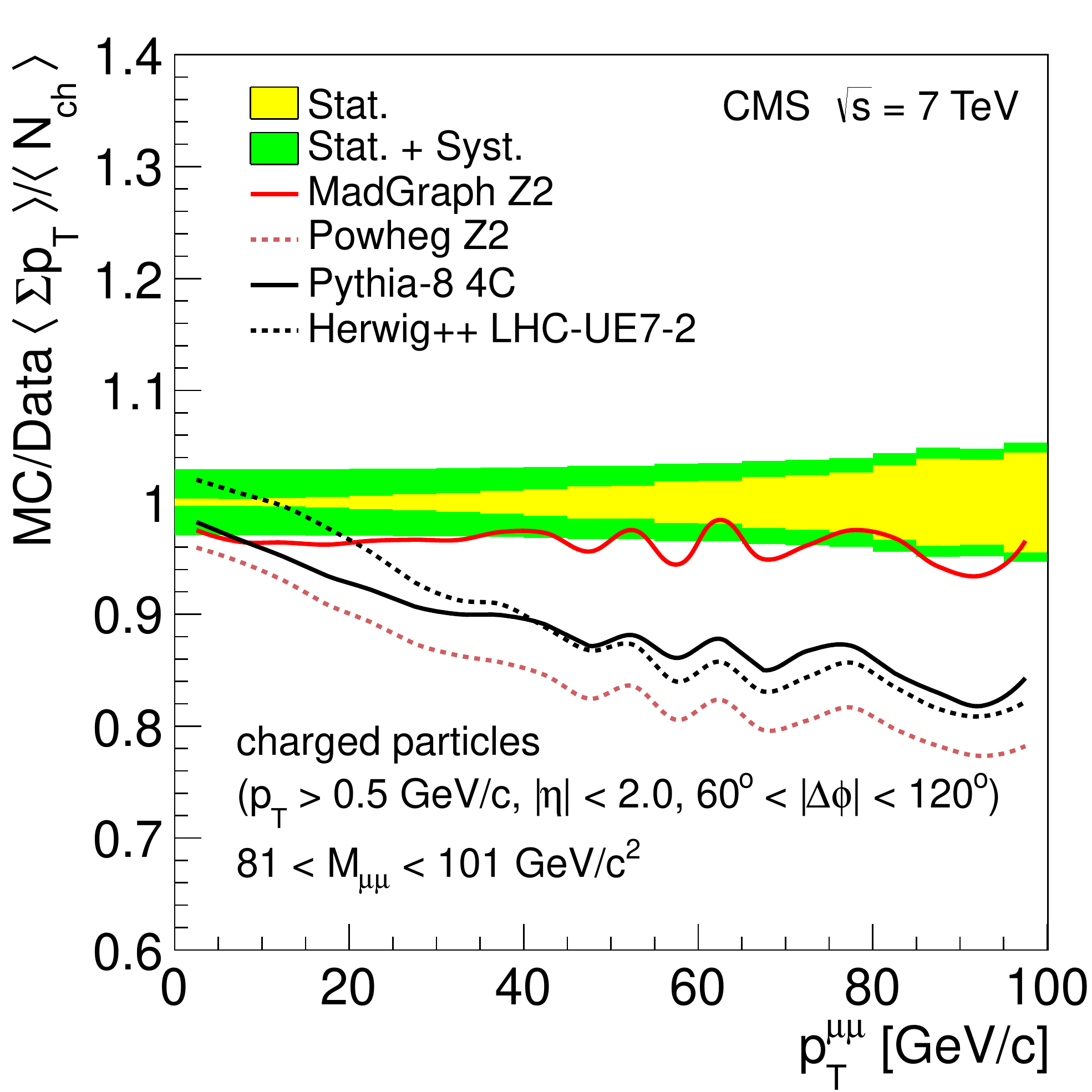}\\
\includegraphics[width=0.325\textwidth]{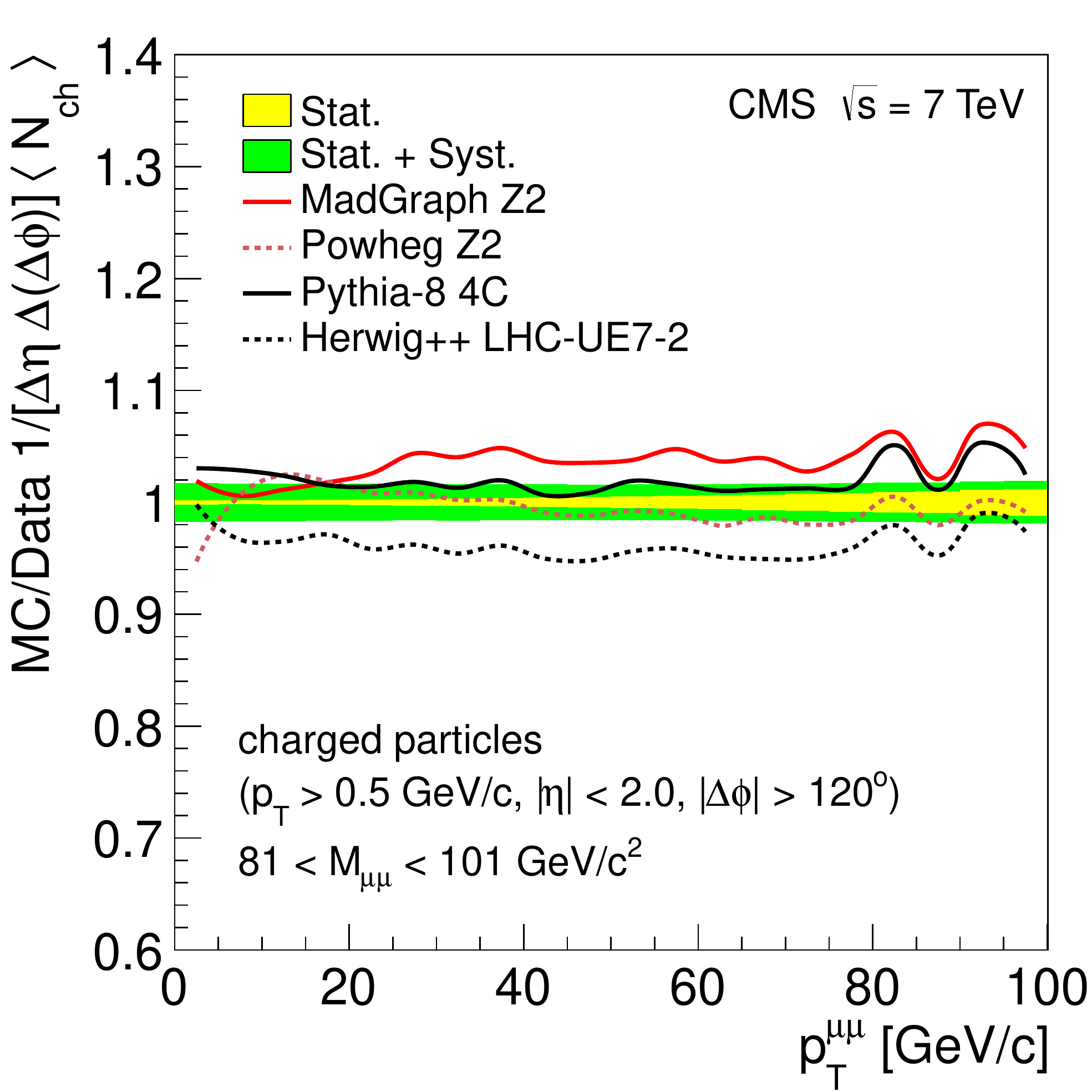} & \includegraphics[width=0.325\textwidth]{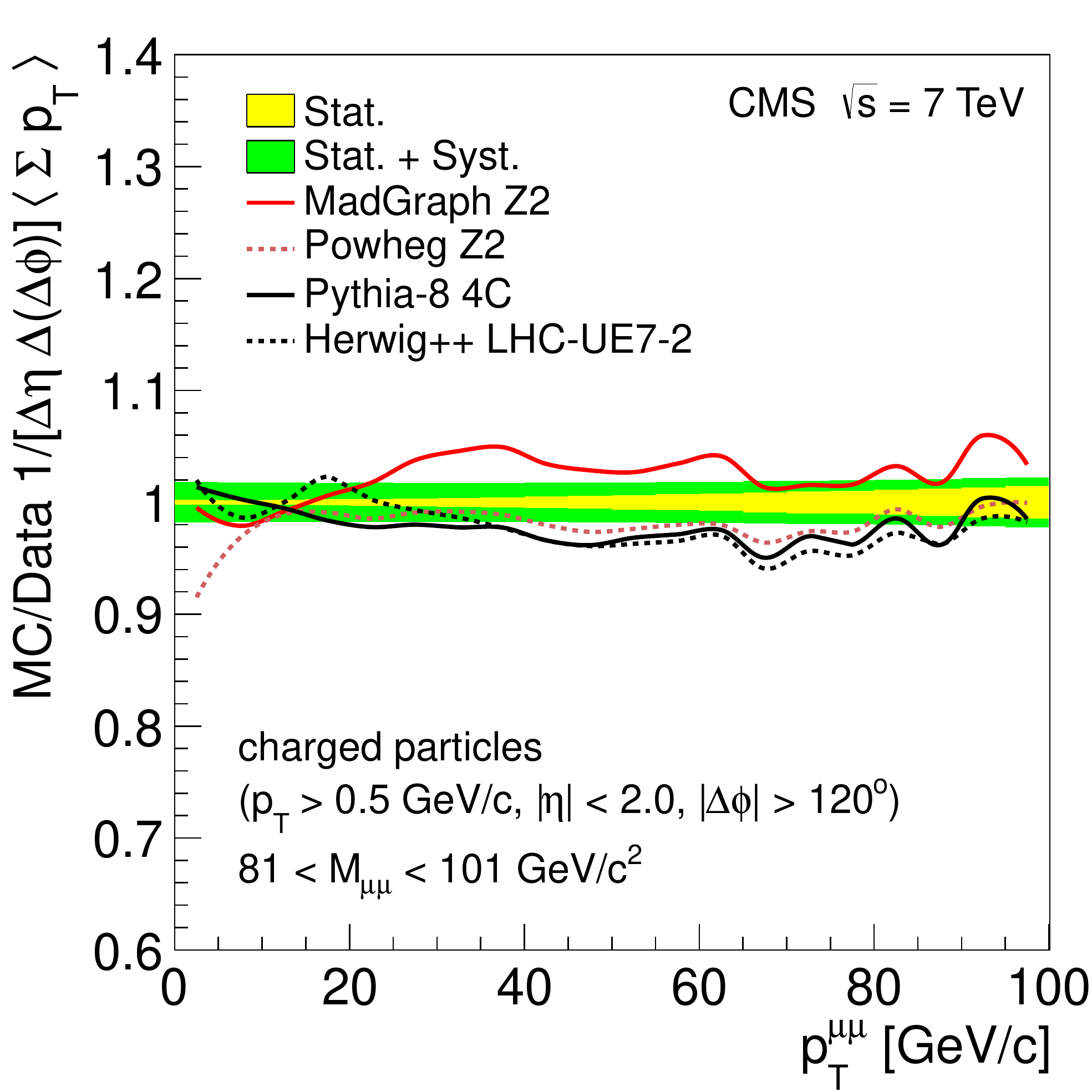} & \includegraphics[width=0.325\textwidth]{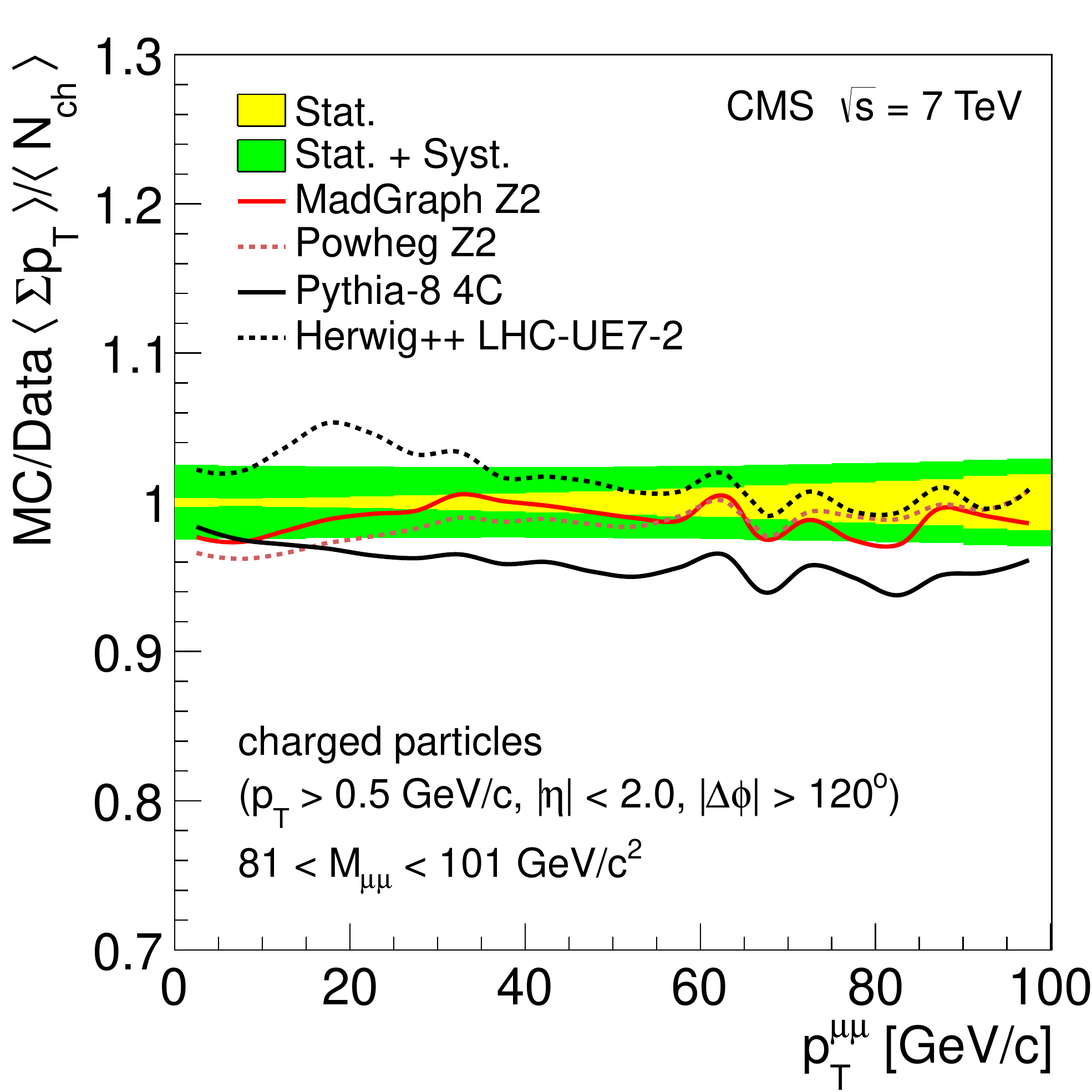}\\
\end{tabular}
 \caption{ Ratios, as functions of $p_{T}^{\mu\mu}$, of the predictions of various MC models to the measurements in the towards (upper row),  transverse (centre row), and away (bottom row) regions for events satisfying $81 < M_{\mu\mu} < 101$\GeVcc: (left) particle density; (centre) energy density; (right) the ratio of the energy density and particle densities. The inner band shows the statistical uncertainty on the data whereas the outer band represents the total uncertainty.}
\label{fig:ratio_profile_pt}
\end{figure*}

Figure~\ref{fig:ratio_profile_pt} presents the ratios of the predictions of various MC models to the measurements for the observables shown in Fig.~\ref{fig:profile_pt}.
 Statistical fluctuations in the data induce correlated fluctuations for the various MC/data ratios.
 \MADGRAPH in conjunction with \PYTHIA{}6 tune Z2 describes the $p_{T}^{\mu\mu}$ dependence of the UE activity very well, both qualitatively and quantitatively.
 \PYTHIA{}8 4C and \HERWIG{}++ describe the $p_{T}^{\mu\mu}$ dependence of the particle density within 10--15\%, but fail to describe the energy density.
 \PYTHIA{}8 4C and \HERWIG{}++ agree better with data as $p_{T}^{\mu\mu}$ approaches zero.
 The combination of the Z2 tune with {\POWHEG fails to describe the energy density in the towards and transverse regions,  but gives a reasonable description of the particle density.
 This observation, combined with the information in Fig.~\ref{fig:profile_mll}, indicates that the discrepancies are not necessarily due to a flaw in the UE tune, but to an inadequate description of the multiple hard emissions and the different sets of PDFs used with \POWHEG}.
 At small $p_{T}^{\mu\mu}$ the comparisons with \PYTHIA{}6 Z2 and \POWHEG Z2 are similar to those in Ref.~\cite{Zpt}, where \PYTHIA{}6 gives a good description of the $p_{T}^{\mu\mu}$ spectrum while \POWHEG underestimates the $p_{T}^{\mu\mu}$.

\begin{figure*}[htbp]
\centering
\includegraphics[width=0.32\textwidth]{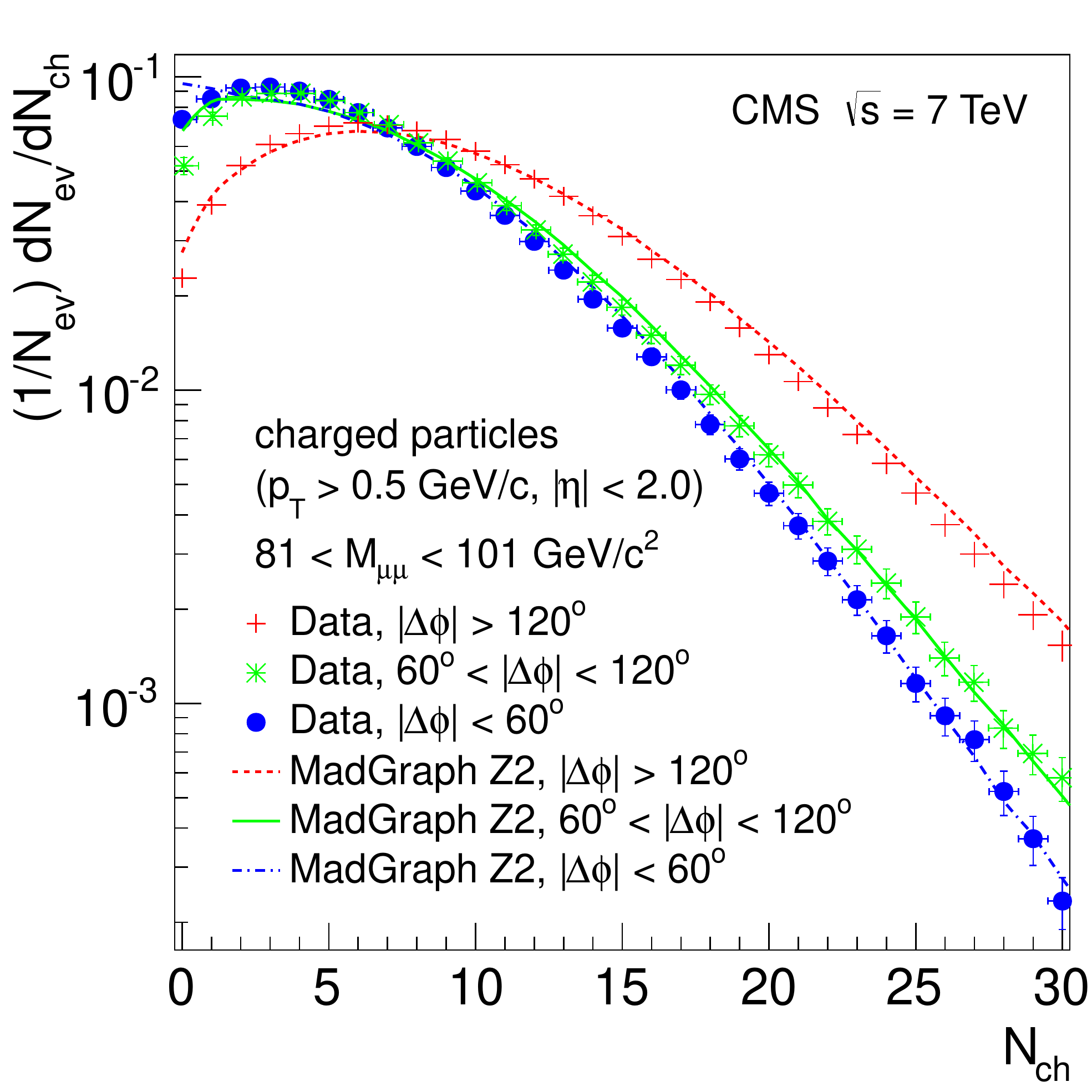}
\includegraphics[width=0.32\textwidth]{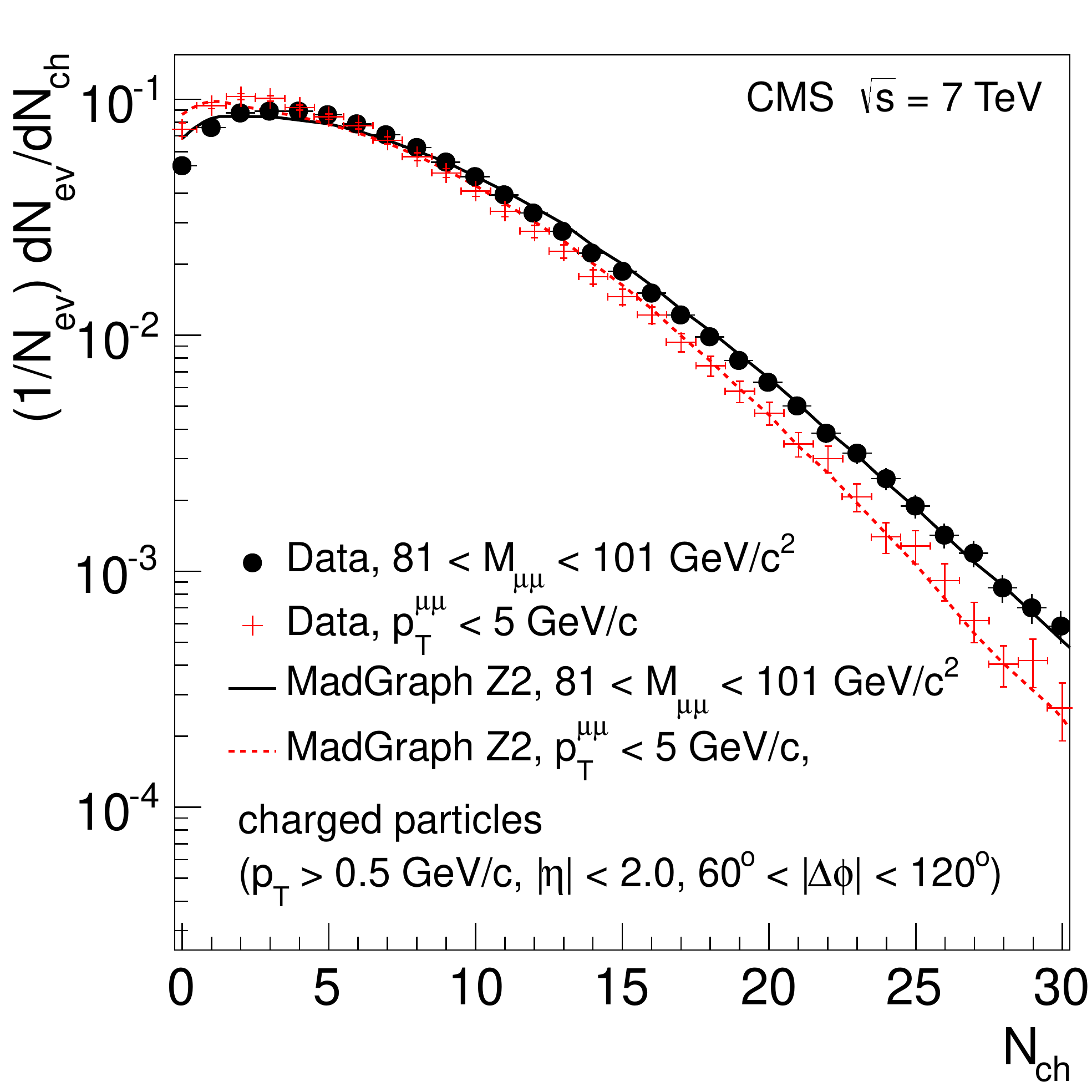}
\includegraphics[width=0.32\textwidth]{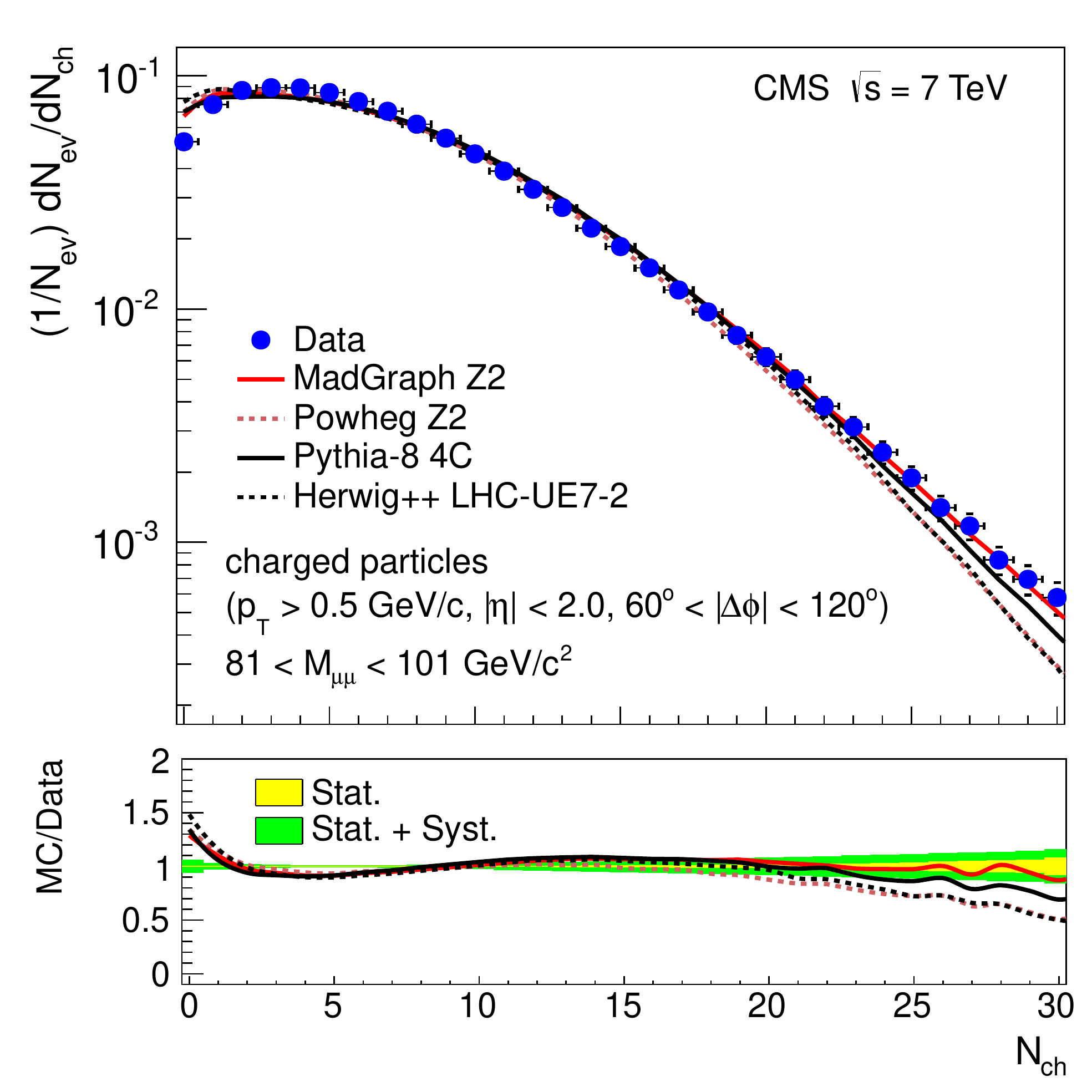}
\includegraphics[width=0.32\textwidth]{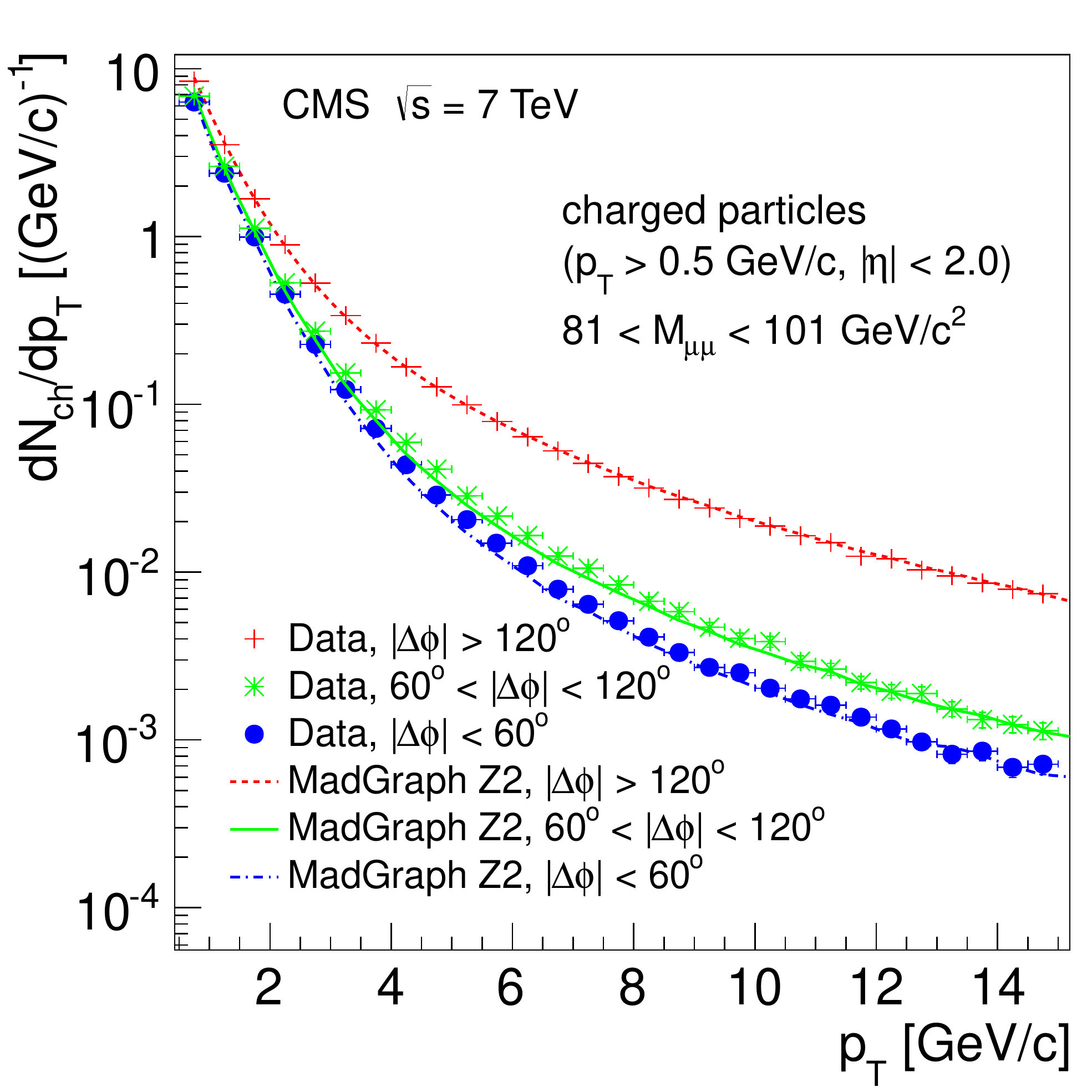}
\includegraphics[width=0.32\textwidth]{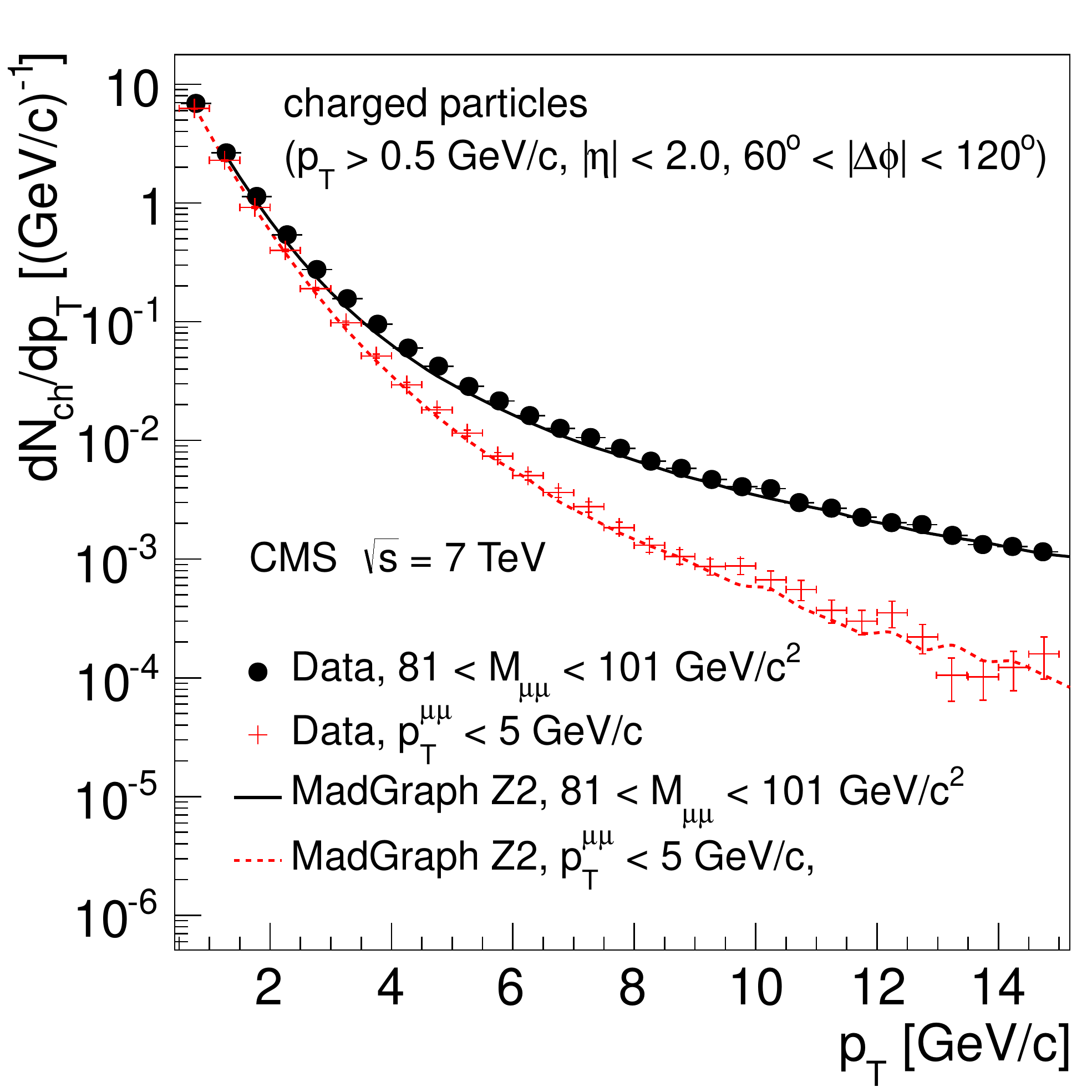}
\includegraphics[width=0.32\textwidth]{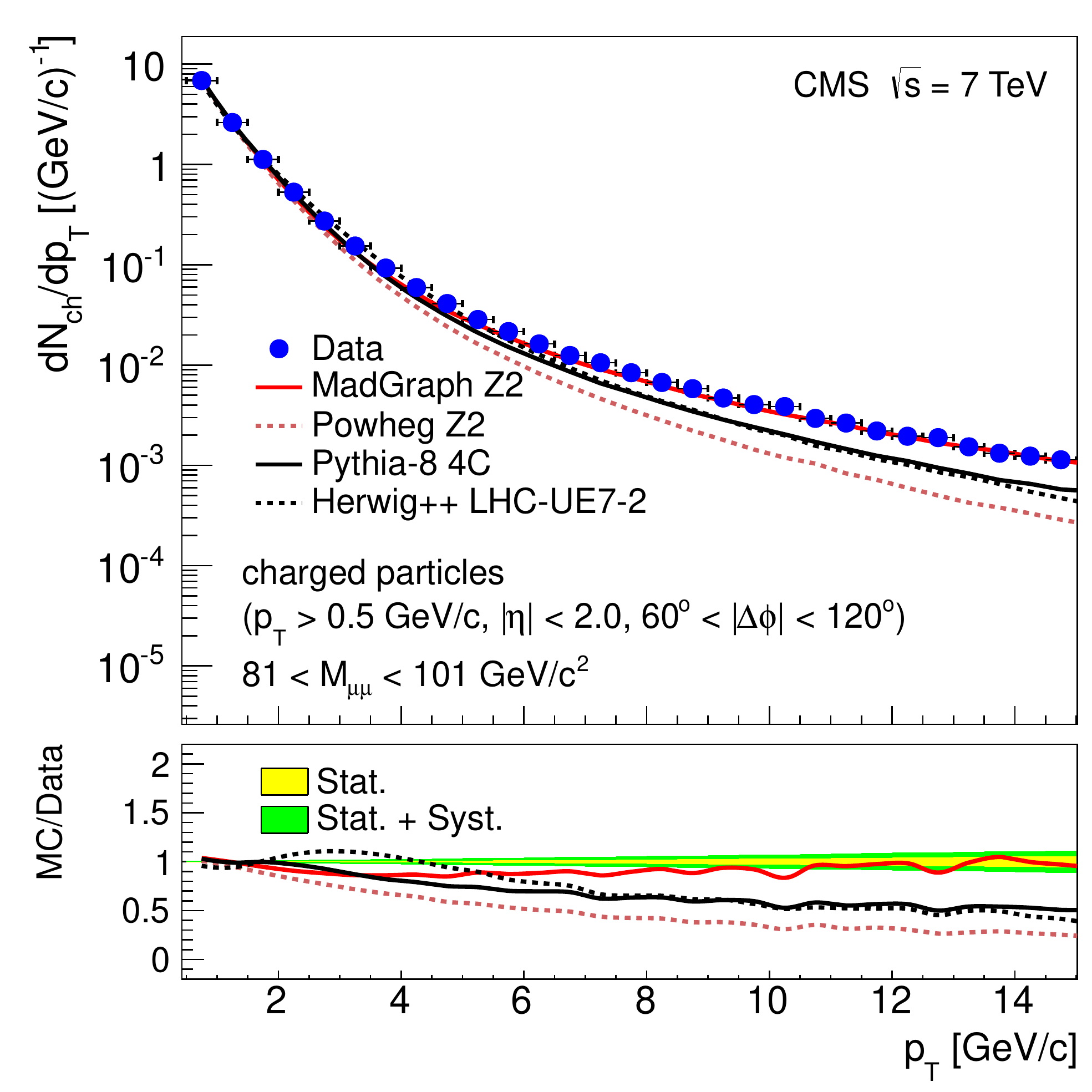}

 \caption{Distributions of the charged particle multiplicity (upper row) and transverse momentum (bottom row) of the selected tracks. The left plots show the comparisons of the normalized distributions in the away, transverse, and towards regions for events satisfying $81 < M_{\mu\mu} < 101$\GeVcc. Comparisons of the normalized distributions in the transverse region are shown in the centre plots, requiring $81 < M_{\mu\mu} < 101$\GeVcc or $p_{T}^{\mu\mu} < 5\GeVc$ . The right plots show the comparisons of the normalized distributions in the transverse region with the predictions of various simulations for events satisfying  $81 < M_{\mu\mu} < 101$\GeVcc.
}
\label{fig:norm}
\end{figure*}

Figure~\ref{fig:norm} shows the distributions of charged particle multiplicity (top row) and transverse momentum (bottom row).
 Figure~\ref{fig:norm} (left) shows a comparison of the normalized distributions in the away, transverse, and towards regions for events satisfying $81 < M_{\mu\mu} < 101$\GeVcc.
 As expected, the transverse and towards regions have fewer charged particles with a softer $p_{T}$ spectrum than the away region.
 Figure~\ref{fig:norm} (centre) shows the comparison of the normalized distributions in the transverse region for two different subsets of the selected events, one with $81 < M_{\mu\mu} < 101\GeVcc$ and one with $p_{T}^{\mu\mu} < 5\GeVc$.
 The charged particle multiplicity is decreased and the $p_{T}$ spectrum is softer when $p_{T}^{\mu\mu} < 5\GeVc$ is required,  because of the reduced contribution of ISR.
 Figure~\ref{fig:norm} (right) shows the comparison of the normalized distributions with the predictions of various simulations in the transverse region for events satisfying $81 < M_{\mu\mu} < 101$\GeVcc.
  The charge multiplicity distribution is described well, within 10--15\%, by \MADGRAPH Z2 and \PYTHIA{}8 4C.
 The $p_{T}$ spectrum is described within 10--15\% by \MADGRAPH Z2, whereas \PYTHIA{}8 4C,
 \POWHEG Z2, and \HERWIG{}++ LHC-UE7-2 have softer $p_{T}$ spectra.
 The various MC programs achieve a similar level of agreement with data in the towards region as in the transverse region.

\subsection{Comparison with the UE activity in hadronic events }
\begin{figure*}[htbp]
\centering
\begin{tabular}{@{}c@{}@{}c@{}@{}c@{}}
\multicolumn{1}{c}{\scriptsize particle density} & \multicolumn{1}{c}{\scriptsize energy density} &  \multicolumn{1}{c}{\scriptsize ratio of energy and particle densities}\\
\includegraphics[width=0.32\textwidth]{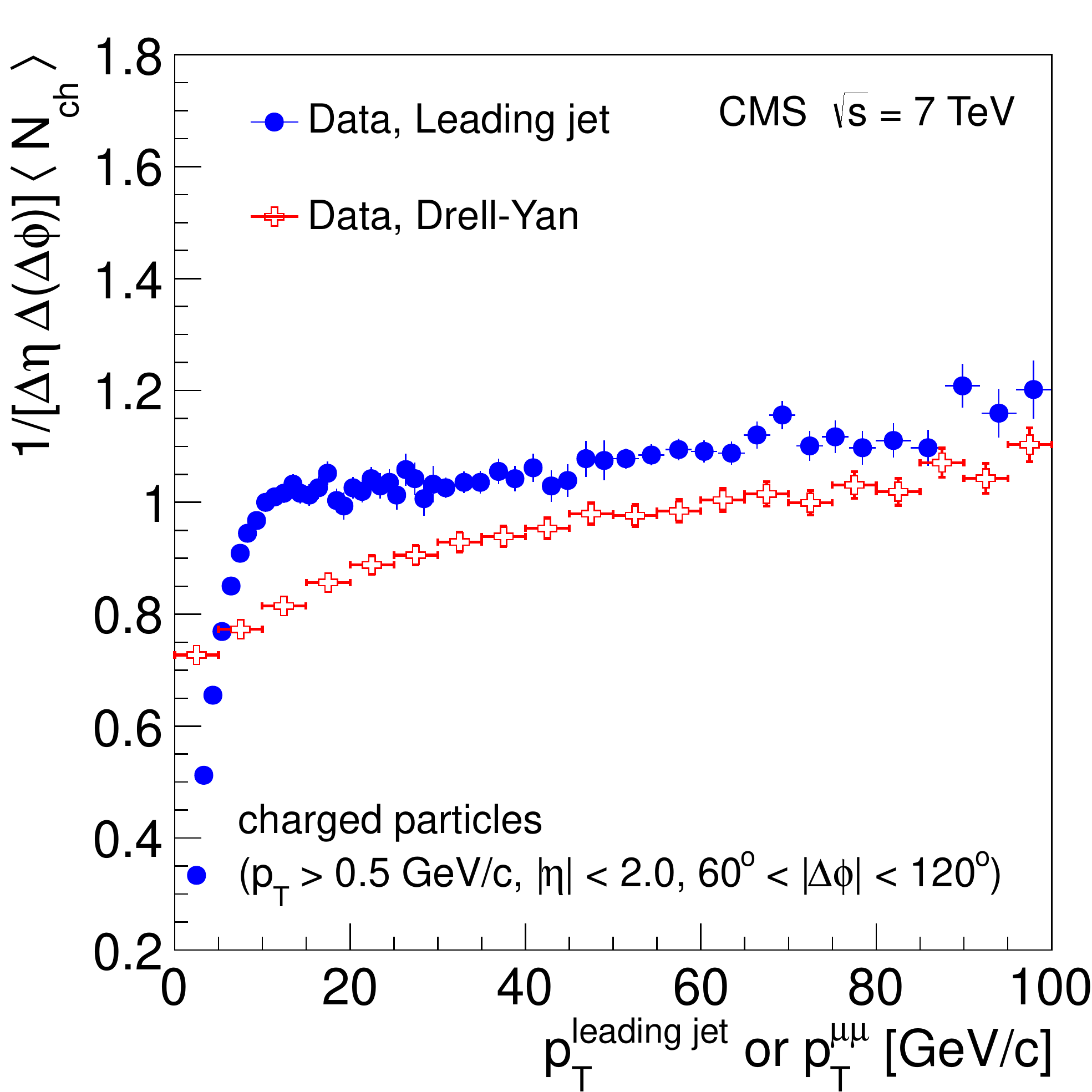} & \includegraphics[width=0.32\textwidth]{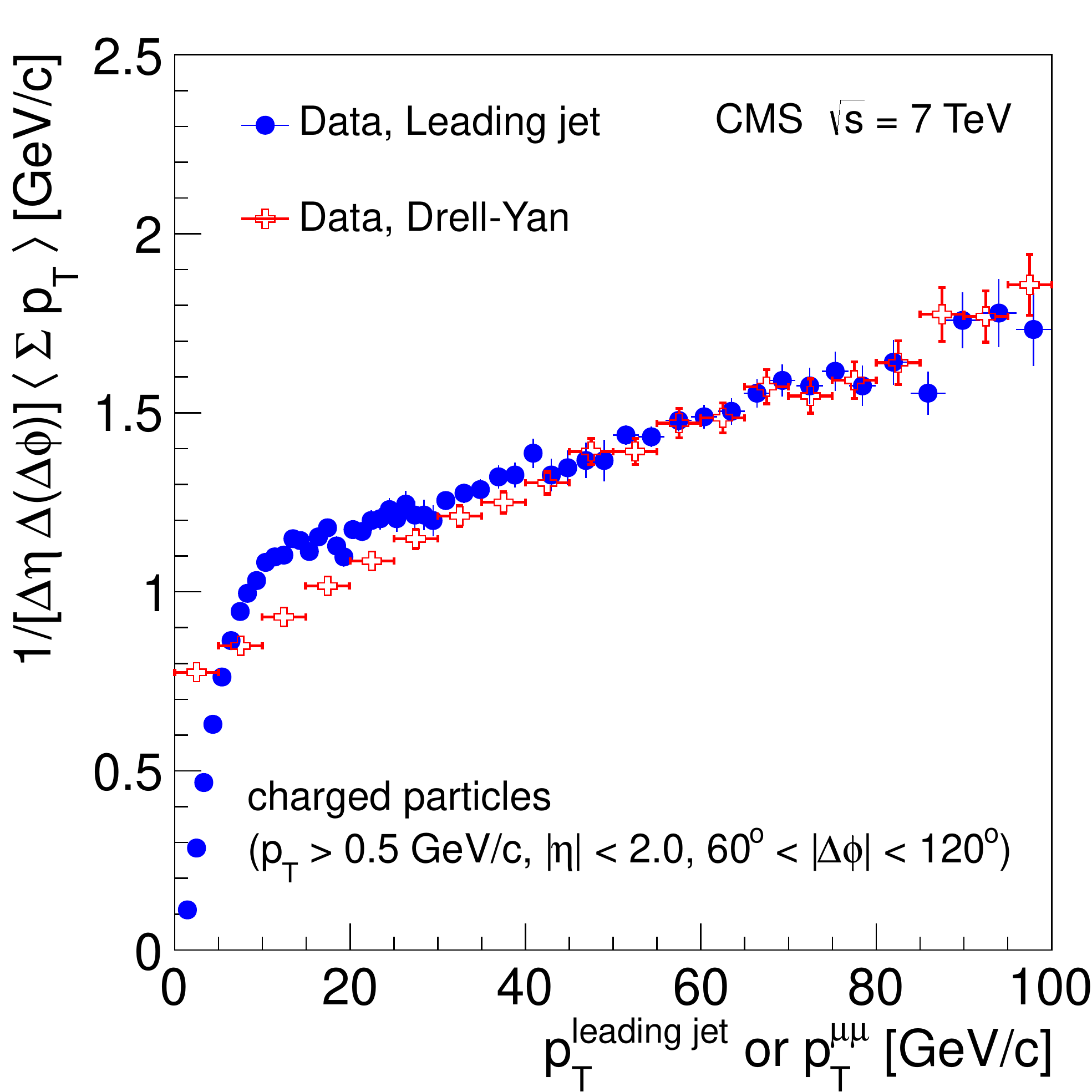} & \includegraphics[width=0.32\textwidth]{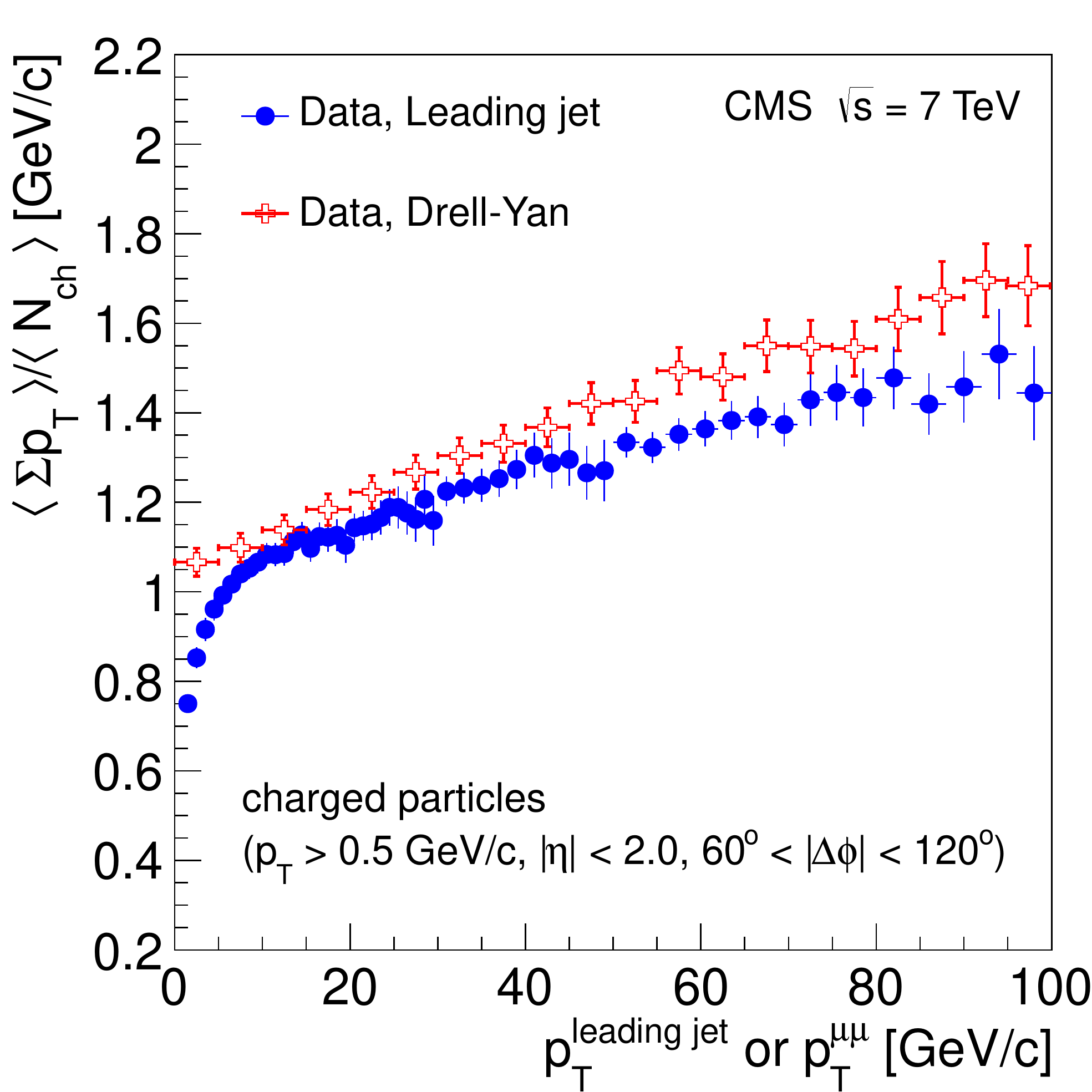}\\
\end{tabular}
 \caption{Comparison of the UE activity measured in hadronic and Drell--Yan events (around the \Z resonance peak) as a function of $p_{T}^\text{leading~jet}$ and $p_{T}^{\mu\mu}$, respectively: (left) particle density, (centre) energy density, and (right) ratio of energy and particle densities in the transverse region.}
\label{fig:trkJet}
\end{figure*}

The UE activity was previously measured as a function of leading jet $p_{T}$ in hadronic events for charged particles with pseudorapidity $|\eta| < 2$ and  with transverse momentum $p_{T} > 0.5$\GeVc~\cite{CMS-PAS-QCD-10-010}.
 Figure~\ref{fig:trkJet} shows the comparison of the UE activity measured in the hadronic and the DY events (around the Z peak) in the transverse region as a function of $p_{T}^{\rm leading~jet}$ and $p_{T}^{\mu\mu}$, respectively.
 For the hadronic events two components are visible: a fast rise for $p_{T}^{\rm leading~jet} \lesssim 10\GeVc$ due to an increase in the MPI activity, followed by an almost constant particle density and a slow increase in the energy density with $p_{T}^{\rm leading~jet}$.
 The increase in the particle and energy densities for $p_{T}^{\rm leading~jet} \gtrsim 10\GeVc$ is mainly due to the increase of ISR and FSR. Owing to the presence of a hard energy scale ($81 < M_{\mu\mu} < 101$\GeVcc), densities in the DY events do not show a sharply rising part, but only a slow growth with $p_{T}^{\mu\mu}$ due to the ISR contribution.

For $p_{T}^{\mu\mu}$ and $p_{T}^{\rm leading~jet} > 10\GeVc$, DY events have a smaller particle density with a harder $p_{T}$ spectrum compared to the hadronic events, as can be seen in Fig.~\ref{fig:trkJet}.
 This distinction is due to the different nature of radiation in the hadronic and DY events.
 Drell--Yan events have only initial-state QCD radiation initiated by quarks, which fragment into a smaller number of hadrons carrying a larger fraction of the parent parton energy, whereas the hadronic events have both initial- and final-state QCD radiation predominantly initiated by gluons with a softer fragmentation into hadrons. Similar behavior is observed for the track-jet measurement where the UE activity is higher by 10--20\% for gluon-dominated processes, as estimated from simulation.

\section{Summary}

We have used Drell--Yan events to measure the UE activity in proton-proton collisions at $\sqrt{s} = 7\TeV$, which were recorded with the CMS detector at the LHC.
 The DY process provides a UE measurement where  a clean separation of the hard interaction from the soft component is possible.
 After excluding the muons from the DY process, the towards ($|\Delta\phi|<60^{\circ}$) and the transverse ($60^{\circ}<|\Delta\phi|<120^{\circ}$) regions are both sensitive to initial-state radiation and multiple parton interactions.
 The DY process provides an effective way to study the dependence of the UE activity on the hard interaction scale, which is related to the invariant mass of the dimuon pair.
 The influence of the ISR is probed by the dependence on the transverse momentum of the muon pair.

The UE activity is observed to be independent of the dimuon mass above 40\GeVcc, after limiting the recoil activity, which confirms the MPI saturation at this scale.
The UE activity in the DY events with no hard ISR is well described by \PYTHIA{}6 and \MADGRAPH with the Z2 tune and the CTEQ6L PDF.
The Z2 tune does not agree with the data if used with PDFs other than CTEQ6L, as in the case of the \POWHEG simulation.
 The \PYTHIA{}8 4C and \HERWIG{}++ LHC-UE7-2 tunes provide good descriptions of the energy-scale dependence of the UE activity.
 Thus the dependence of the UE activity on the energy scale is well described by tunes derived from  hadronic events, illustrating the universality of MPIs in different processes.
 This universality is also indicated by the similarity between the UE activity in DY and hadronic events, although these events have different types of radiation.
 In addition, there is some ambiguity in the definition of the hard scale for both types of events.

The UE activity in the towards and transverse regions shows a slow growth with the transverse momentum of the muon pair and provides an important probe of the ISR.
 The leading-order matrix element generator \MADGRAPH provides a good description of the UE dependence on dimuon transverse momentum. However, \PYTHIA, \POWHEG, and \HERWIG{}++, which do not simulate the multiple hard emissions with sufficient accuracy, underestimate the energy density, but describe the particle density reasonably well.
These measurements provide important input for further tuning or improvements of the Monte Carlo models and also for the understanding of the dynamics of QCD.

\section*{Acknowledgements}
\hyphenation{Bundes-ministerium Forschungs-gemeinschaft Forschungs-zentren}
We wish to congratulate our colleagues in the CERN accelerator departments for the excellent performance of the LHC machine. We thank the technical and administrative staff at CERN and other CMS institutes. This work was supported by the Austrian Federal Ministry of Science and Research; the Belgium Fonds de la Recherche Scientifique, and Fonds voor Wetenschappelijk Onderzoek; the Brazilian Funding Agencies (CNPq, CAPES, FAPERJ, and FAPESP); the Bulgarian Ministry of Education and Science; CERN; the Chinese Academy of Sciences, Ministry of Science and Technology, and National Natural Science Foundation of China; the Colombian Funding Agency (COLCIENCIAS); the Croatian Ministry of Science, Education and Sport; the Research Promotion Foundation, Cyprus; the Estonian Academy of Sciences and NICPB; the Academy of Finland, Finnish Ministry of Education and Culture, and Helsinki Institute of Physics; the Institut National de Physique Nucl\'eaire et de Physique des Particules~/~CNRS, and Commissariat \`a l'\'Energie Atomique et aux \'Energies Alternatives~/~CEA, France; the Bundesministerium f\"ur Bildung und Forschung, Deutsche Forschungsgemeinschaft, and Helmholtz-Gemeinschaft Deutscher Forschungszentren, Germany; the General Secretariat for Research and Technology, Greece; the National Scientific Research Foundation, and National Office for Research and Technology, Hungary; the Department of Atomic Energy and the Department of Science and Technology, India; the Institute for Studies in Theoretical Physics and Mathematics, Iran; the Science Foundation, Ireland; the Istituto Nazionale di Fisica Nucleare, Italy; the Korean Ministry of Education, Science and Technology and the World Class University program of NRF, Korea; the Lithuanian Academy of Sciences; the Mexican Funding Agencies (CINVESTAV, CONACYT, SEP, and UASLP-FAI); the Ministry of Science and Innovation, New Zealand; the Pakistan Atomic Energy Commission; the State Commission for Scientific Research, Poland; the Funda\c{c}\~ao para a Ci\^encia e a Tecnologia, Portugal; JINR (Armenia, Belarus, Georgia, Ukraine, Uzbekistan); the Ministry of Science and Technologies of the Russian Federation, the Russian Ministry of Atomic Energy and the Russian Foundation for Basic Research; the Ministry of Science and Technological Development of Serbia; the Ministerio de Ciencia e Innovaci\'on, and Programa Consolider-Ingenio 2010, Spain; the Swiss Funding Agencies (ETH Board, ETH Zurich, PSI, SNF, UniZH, Canton Zurich, and SER); the National Science Council, Taipei; the Scientific and Technical Research Council of Turkey, and Turkish Atomic Energy Authority; the Science and Technology Facilities Council, UK; the US Department of Energy, and the US National Science Foundation.
Individuals have received support from the Marie-Curie programme and the European Research Council (European Union); the Leventis Foundation; the A. P. Sloan Foundation; the Alexander von Humboldt Foundation; the Belgian Federal Science Policy Office; the Fonds pour la Formation \`a la Recherche dans l'Industrie et dans l'Agriculture (FRIA-Belgium); the Agentschap voor Innovatie door Wetenschap en Technologie (IWT-Belgium); and the Council of Science and Industrial Research, India.

\clearpage

\bibliography{auto_generated}

\cleardoublepage \appendix\section{The CMS Collaboration \label{app:collab}}\begin{sloppypar}\hyphenpenalty=5000\widowpenalty=500\clubpenalty=5000\textbf{Yerevan Physics Institute,  Yerevan,  Armenia}\\*[0pt]
S.~Chatrchyan, V.~Khachatryan, A.M.~Sirunyan, A.~Tumasyan
\vskip\cmsinstskip
\textbf{Institut f\"{u}r Hochenergiephysik der OeAW,  Wien,  Austria}\\*[0pt]
W.~Adam, T.~Bergauer, M.~Dragicevic, J.~Er\"{o}, C.~Fabjan, M.~Friedl, R.~Fr\"{u}hwirth, V.M.~Ghete, J.~Hammer\cmsAuthorMark{1}, M.~Hoch, N.~H\"{o}rmann, J.~Hrubec, M.~Jeitler, W.~Kiesenhofer, M.~Krammer, D.~Liko, I.~Mikulec, M.~Pernicka$^{\textrm{\dag}}$, B.~Rahbaran, C.~Rohringer, H.~Rohringer, R.~Sch\"{o}fbeck, J.~Strauss, A.~Taurok, F.~Teischinger, P.~Wagner, W.~Waltenberger, G.~Walzel, E.~Widl, C.-E.~Wulz
% ...
\vskip\cmsinstskip
\dag:~Deceased\\
1:~~Also at CERN, European Organization for Nuclear Research, Geneva, Switzerland\\
% ...

\end{sloppypar}
\end{document}